\documentclass{ar-1col}
\pdfoutput=1

\usepackage{rotating}
\renewcommand{\r}{{\bf r}}
\jname{Xxxx. Xxx. Xxx. Xxx.}
\jvol{00}
\jyear{YYYY}
\doi{10.1146/((please add article doi))}

\newcommand{\eqref}[1]{Eq.~\ref{#1}}
\renewcommand{\r}{{\bf r}}
\renewcommand{\u}{{\bf u}}
\newcommand{\Pe}{{\hbox{\rm Pe}}}

\begin{document}

% Page heads
\markboth{Michael Cates and Julien Tailleur}{Motility-Induced Phase Separation}

% Title
\title{Motility-Induced Phase Separation}

% Author/affiliation
\author{Michael E. Cates$^1$ and Julien Tailleur$^2$
\affil{$^1$SUPA, School of Physics and Astronomy, University of Edinburgh, JCMB Kings Buildings, Mayfield Road, Edinburgh EH9 3JZ, UK; m.e.cates@ed.ac.uk}
\affil{$^2$Univ Paris Diderot, Sorbonne Paris Cit\'e, MSC, UMR 7057 CNRS, F75205 Paris, France; julien.tailleur@univ-paris-diderot.fr}}

% Abstract
\begin{abstract}
Self-propelled particles include both self-phoretic synthetic 
colloids and various micro-organisms. By continually
consuming energy, they bypass the laws of
equilibrium thermodynamics. These laws enforce the Boltzmann
distribution in thermal equilibrium: the steady state is then
independent of kinetic parameters. In contrast,
self-propelled particles tend to accumulate where they move more
slowly. They may also slow down at high density, for either 
biochemical or steric reasons. This creates positive feedback which can 
lead to
motility-induced phase separation (MIPS) between dense and dilute fluid
phases. At leading order in gradients, a mapping relates variable-speed,
self-propelled particles to passive particles with attractions. 
This deep link to equilibrium phase separation is
confirmed by simulations, but generally breaks down at higher
order in gradients: new effects, with no equilibrium counterpart, then emerge. We give a selective overview of the fast-developing
field of MIPS, focusing on theory and simulation but including a brief
speculative survey of its experimental implications.
\end{abstract}

% Keywords

\maketitle

% to generate article TOC
\tableofcontents

% Head 1
\section{INTRODUCTION}

Non-equilibrium systems arise in a wide range of situations with very
different phenomenologies. Nonetheless, one can identify general
categories that share sufficient ingredients to form coherent
classes. One such class describes systems that are relaxing towards,
but have not yet reached, thermal equilibrium. This relaxation may be
relatively unhindered, or might become extremely slow (as happens in
glasses). Nonetheless, there is a sense of a direction in which the
system either moves, or would move if it could. A second class of
non-equilibrium systems describes those whose bulk dynamics is
prevented from attaining equilibrium by boundary conditions imposing
non-zero steady currents. These are exemplified by heat flow
experiments, in which a piece of matter is connected to two reservoirs
held at different temperatures.

In a third class of non-equilibrium systems, often called `active
matter', energy is dissipated at the microscopic scale \textit{in the
  bulk}, so that each constituent of the system has an irreversible
dynamics. This includes a large range of systems whose particles are
`motile', i.e., self-propelled: bird flocks~\cite{PNASGiardina}, fish
schools~\cite{poisson}, actin filaments~\cite{Bausch} and
microtubules~\cite{Sumino} in motility assays, autophoretic
colloids~\cite{TheurkauffPRL,ButtinoniBechinger,PalacciScience}, and
``colloidal rollers"~\cite{DB}. Many such systems have been studied in
their own right~\cite{Vicsek}, but over the past ten years, the quest
for a generic description of active matter has attracted growing
interest~\cite{ABP_review,CatesRPP,ASMRMP}. It is reasonable to hope
that self-propelled particles, which otherwise interact via standard
equilibrium forces (attraction, repulsion, alignment, etc.), might
form a coherent sub-class of non-equilibrium systems that can be
described by a common theoretical framework.

Active matter systems can exhibit many new behaviors, at least some of
which should prove relevant to applications. For instance, many forms
of bacterial contamination (from coronary implants to domestic water
systems \cite{fouling,valves}) arise from biofilm formation on
surfaces, whose early stages are triggered when the local population
density exceeds some threshold~\cite{biofilms}. It would therefore be
helpful to understand how a nonuniform density can arise from a
uniform one.  Also, many studies of motile synthetic colloids (for
instance \cite{PalacciScience}) are motivated by a desire to direct
the assembly of nanostructures. More generally one would like to
extend our control over soft matter systems, whose applications range
from liquid crystal displays to cosmetics and food processing
industries, to include `hybrid' materials in which at least some of
the components are active. Examples might include tissue scaffolds
\cite{scaffolds} for wound repair and electrode systems for microbial
fuel cells \cite{MFCs}.

The study of active matter has historically been driven mainly by work
on biological systems. These are, however, often quite complex: for
instance, the rich phenomenology observed in dense swarms of bacteria
stems in combination from their self-propulsion, the alignment
interactions due to their rod-like shapes, and their hydrodynamic
coupling to the medium in which they swarm~\cite{goldstein2}. One
fruitful line of research, following the seminal work of Vicsek and
co-workers~\cite{Vicsek}, has been dedicated to motile particles with
orientational order. This spans polar particles~\cite{Vicsek}, active
nematics~\cite{activenema}, self-propelled rods~\cite{Peruani1} and
active Ising spins~\cite{activeising}; the interplay between
interaction-induced alignment of motile particles and their
self-propulsion has led to the discovery of a variety of new phases
(such as the `zooming bionematic' phase \cite{cisneros}) and various
transitions between these phases.

Inspired by the increasing experimental availability of synthetic
motile colloids of somewhat simpler geometry, theorists have recently
also addressed simpler models in which (because they are dilute or of
spherical shape) swimmers have no innate tendency towards
orientational order. Even dilute active suspensions can give
nontrivial density profiles in sedimentation
equilibrium~\cite{JTEPL,Palacci2010,StarkPRL}, and their microscopic
irreversibility becomes manifest via rectification interactions with
mesoscopic ratchets~\cite{Austin1,wan,JTEPL,rotorpaper1,rotorpaper2,rotorpaper3}. In these dilute cases interactions between particles can essentially be neglected,
showing that the active equivalent of an ideal gas is already a highly
non-trivial object of study.

A first step beyond this non-interacting limit is to consider `active
simple fluids' made of spherical self-propelled particles whose
interactions are isotropic. The latter not only includes conventional
attractive and repulsive interactions of colloidal spheres but also allows for
some forms of signaling, such as quorum sensing in bacteria
\cite{quorum}, where a chemical species is emitted isotropically by
each particle and its concentration detected. (Note that bacteria
cannot directly detect vector quantities such as chemical
concentration gradients, but do so indirectly by integrating temporal
information as they move. This means they lack the long-range
orientational interactions of, say, bird flocks, in which individuals
visually detect the mean velocity vector of their neighbors.) Even
without any orientational interactions, active simple fluids have a
phenomenology much richer than their passive counterparts. Most
notably, self-propelled particles with purely repulsive interactions
can undergo liquid-gas phase
separation~\cite{thompson,Fily,RednerPRL,BialkeLowen,Ludo}. This is impossible
for passive colloidal particles without attractions and stems from an
intrinsically non-equilibrium mechanism, called Motility-Induced Phase
Separation (MIPS)~\cite{TCPRL}, which is the subject of this
review. When the speed of motile particles decreases sufficiently
steeply as their local density increases, a uniform suspension becomes
unstable, leading to a phase-separated state where a dilute active gas
coexists with a dense liquid of substantially reduced
motility.\begin{marginnote} \entry{MIPS}{Motility-Induced Phase
    Separation}
\end{marginnote}

Before reviewing this surprisingly generic phenomenon in depth, we
summarize its underlying mechanism, which can be intuitively captured
by a relatively simple argument. A first ingredient, carefully
explored by Schnitzer for the case of run-and-tumble
bacteria~\cite{SchnitzerPRE}, is that active particles generically
accumulate where they move more slowly. This follows directly from the
master equation of a self-propelled particle of spatially varying
speed $v(\r)$:
\begin{equation}
  \label{eqn:ME}
  \dot P(\r,\u) = - \nabla \cdot [v(\r)\u P(\r,\u) ] + \Theta[P(\r,\u)]
\end{equation}
Here $\Theta[P]$ accounts for the changes of the particle orientation
$\u$. (For instance, $\Theta[P]=D_r \Delta_\u P$ for Brownian
rotational diffusion.) For isotropic processes, $P_{\rm
  stat}(\r,\u)\propto 1/v(\r)$ is always a steady-state solution
of~\eqref{eqn:ME}. This effect is 
crucially absent for the Brownian motion of particles in thermal
equilibrium. In that case, $v$ is a random variable whose statistics
are entirely independent of $\r$, instead depending solely on
temperature. (This is the equipartition theorem for kinetic energies.)
Therefore a similar effect can only arise in a passive system if the temperature is
nonuniform~\cite{VK,SchnitzerBerg}: although the diffusivity of
isothermal Brownian particles might vary with position, for instance
due to gradients of viscosity, this has no effect on $P_{\rm stat}$,
which at uniform temperature is a function of energy only.

The second crucial ingredient of MIPS arises in an assembly of active
particles exhibiting a propulsion speed $v$ that depends on the local
particle density $\rho$.  Such a dependence might arise directly by
chemical signaling (e.g., quorum sensing \cite{quorum}) or by
coarse-graining a traditional colloidal interaction such as steric
exclusion (see Sec.~\ref{ABPs} below). MIPS arises from the positive
feedback between this accumulation-induced slowing and the
slowing-induced accumulation implicit in~\eqref{eqn:ME}.
Heuristically, consider a small perturbation $\delta \rho(\r)$ around
a uniform profile $\rho_0\equiv c/v(\rho_0)$ where $c$ is some
constant.  This leads to a spatially varying speed
$v(\rho_0+\delta\rho(\r))=v(\rho_0)+v'(\rho_0) \delta \rho(\r)$ so
that $\delta \rho(\r)$ and $\delta v(\r)$ are in antiphase if
$v(\rho)$ decreases with $\rho$. The steady state density for this
$v(\r)$ would be $\rho_0+\delta\rho'$ where
\begin{equation}
  \rho_0+\delta\rho'=\frac{c}{v(\rho_0)+v'(\rho_0)\delta \rho}\simeq \frac{c}{v(\rho_0)} \left(1-\frac{v'(\rho_0)}{v(\rho_0)}\delta\rho\right)=\rho_0-\rho_0 \frac{v'(\rho_0)}{v(\rho_0)}\delta\rho
\end{equation}
A linear instability is therefore expected whenever $\delta\rho'>\delta\rho$, i.e., when 
\begin{equation}
\label{eqn:InstabCriterion}
\frac{v'(\rho_0)}{v(\rho_0)}< - \frac{1}{\rho_0}
\end{equation}

As we will show below,~\eqref{eqn:InstabCriterion} correctly identifies the region where macroscopic MIPS is initiated by spinodal decomposition. 
One can however go far beyond this
simple linear stability analysis. Indeed, in large measure the coarse-grained
dynamics of active simple fluids can be mapped onto the equilibrium
dynamics of \textit{passive} simple fluids with attractive interactions. This allows a large body of knowledge on that case to be easily transferred. However, the departures from this mapping are also interesting, since they point to aspects where the underlying non-equilibrium character of MIPS cannot be transformed away.

This underlying character stems directly from the lack of microscopic time-reversal symmetry in active systems, which means that their steady states need not obey the principle of detailed balance (DB).
\begin{marginnote} \entry{DB}{Detailed Balance}
\end{marginnote}
This principle states that if phase space is divided up into regions,
the probability flux from region $A$ into region $B$ is the same as
the reverse flux. This precludes circulating fluxes in steady state
(for instance $A\rightarrow B\rightarrow C\rightarrow A$). Such fluxes
are commonly seen in boundary driven systems (such as B\'enard
convection rolls) and can also feature prominently in active
matter~\cite{CatesRPP,JTEPL,rotorpaper1,rotorpaper2,rotorpaper3}. In active matter
DB violations are always present microscopically, but may or may not
survive coarse-graining. Only when they do not survive, is any mapping
possible onto an equivalent equilibrium system.

In what follows we review (Sec.~\ref{Sec2}) the general physics of
motile particles, focusing on two simple models, inspired respectively
by bacteria and by synthetic colloidal swimmers. We then address their
many body physics in general (Sec.~\ref{Sec3}), and MIPS in
particular, first within a local approximation whereby the swim speed
depends on density but not its gradients (Sec.~\ref{Sec4}).  After
exploring the aspects of MIPS for which this approximation is
sufficient (Secs. \ref{Sec5},\ref{Sec7}) we move beyond it, showing in
Sec.~\ref{Sec8} that a careful consideration of nonlocal or gradient
terms gives dynamics that is, after all, not equivalent to any form of
passive phase separation. We conclude briefly in Sec.~\ref{Sec9}.

\section{MOTILE PARTICLES}
\label{Sec2}
% Head 2
\subsection{Run-and-Tumble Bacteria and Active Brownian Particles}

We start by considering two limiting models of the stochastic dynamics
of a single active particle (Fig.1). The first is a so-called
`run-and-tumble particle' (RTP), whose motion consists of periods of
persisent swimming motion, called `runs', punctuated by sudden changes
of direction, called `tumbles' \cite{SchnitzerPRE,BergBook}. This is a
canonically simplified model of the dynamics of bacteria such as {\em
  E. coli}. It supposes the runs to be straight lines, traversed with
fixed speed $v$, and punctuated at random by instantaneous tumbles,
occurring at some fixed rate $\alpha$, each of which completely
decorrelates the swimming direction. At time- and length-scales much
larger than $\alpha^{-1}$ and $\ell \sim v/\alpha$, 
this motion is a diffusive random
walk. 
\footnote{ As so far described, this model is nothing but the Lorentz
  gas introduced to model electron transport in metals. Under the
  assumption that background atoms are random immobile scatterers and
  that the electron-atom interactions amount to elastic scattering,
  electrons indeed undergo run-and-tumble motion~\cite{Krapivsky}.}
It is a simple exercise to calculate its
diffusivity %~\footnote{{Suppressing the fluctuations of run
%    lengths, by considering a process in which tumbles occur every
%    $\Delta t=\alpha^{-1}$, halves the diffusivity to yield
%    $D=v^2/2d\alpha$.}}  
in $d$ dimensions as $D = v^2/\alpha d$.

\begin{marginnote}
\entry{RTP}{Run-and-Tumble Particle}
\entry{ABP}{Active Brownian Particle}
\entry{PBP}{Passive Brownian Particle}
\end{marginnote}

\begin{figure}
  \centering\includegraphics[width=4.75in]{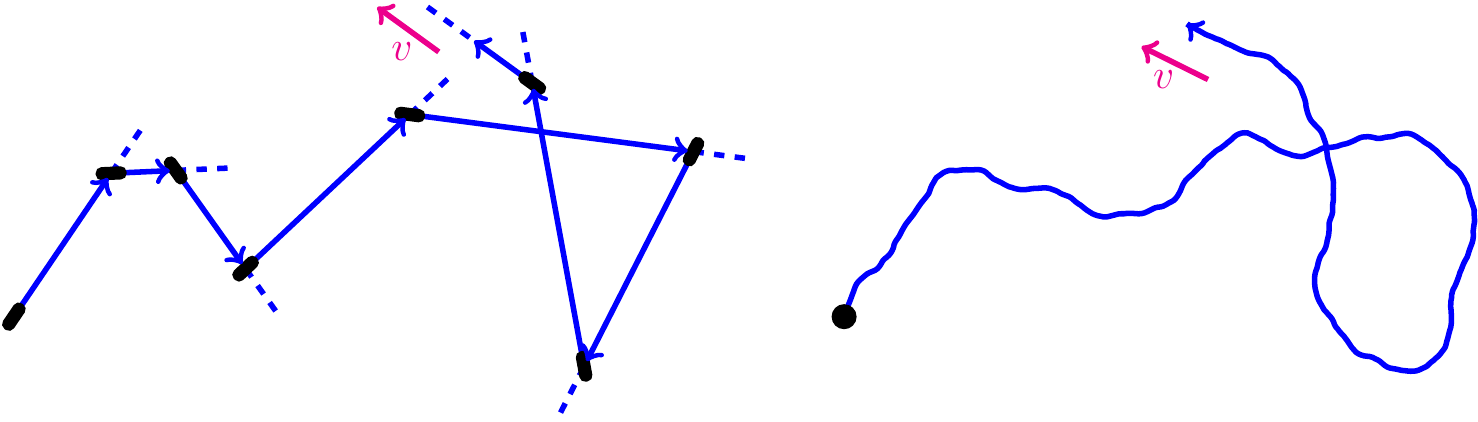}
\caption{{Simulated paths of a run-and-tumble (Left) and an active
  Brownian (Right) particles of length
  $\tau=5 \alpha^{-1}=5 D_r^{-1}$.} Each is diffusive at large length and
  time scales.}\label{fig:one}
\end{figure}

Our second model is called an active Brownian particle (ABP)
\cite{ABP_review}. This also has a fixed swim speed $v$, but its
direction decorrelates smoothly via rotational diffusion, with angular
diffusivity $D_r$. This rotation is typically thermal, hence the
`Brownian' label: an instance is self-phoretic colloids, which
asymmetrically catalyze conversion of a surrounding fuel to create
self propulsion along an axis that slowly rotates by angular Brownian
motion. (Another instance is {\em E. coli} mutants, called `smooth
swimmers', which have lost the ability to tumble.)  At large length-
and time-scales the motion is again a diffusive random walk; finding the
diffusivity is another easy exercise with the result $D =
v^2/d(d-1)D_r$.

As shown in \cite{EPL2}, we can generalize these two calculations of
the large scale diffusivity $D$ to include a superposition of the ABP
and RTP dynamics, and also to include a purely thermal direct
contribution $D_t$ to the translational diffusivity. The result is
\begin{eqnarray}
D &=& \frac{v^2\tau}{d} + D_t \\
\tau^{-1} &= &\alpha + (d-1) D_r
\end{eqnarray}
where $\tau$ is the orientational relaxation time of the
active particle. Note that the $D_t$ contribution is often negligible
compared to the active part; it is frequently set to zero in simulation
studies, and will sometimes be silently omitted in what follows.

\subsection{Spatial Variations in Motility Parameters: a mesoscopic approach}
The above results establish a {\em prima-facie} connection between a
broad generic class of active particle dynamics (with RTP and ABP as
limiting cases) and the physics of isothermal passive Brownian
particles (PBPs): after suitable coarse-graining, all describe diffusive random
walks of the type canonically exemplified by PBPs.
At first sight, the effect of activity is simply to increase the
diffusivity $D$ from that of the equivalent passive particle, typically by a large factor. However, a subtler aspect of the connection to PBPs is revealed if one allows
$\tau$ and $v$ to be functions of the
particle's position, ${\bf r}$ \cite{SchnitzerPRE,SchnitzerBerg}. For instance, a nontrivial $v(\r)$ would arise for bacteria swimming in a polymer gel of variable strand density, so propulsion is more effective in some regions than others.
An
explicit coarse-graining~\cite{EPL2} of the microscopic dynamics
 gives in this case the following equations for the
probability density $\varphi(\r)$ of our single active particle and
its flux~${\bf j}$:
\begin{eqnarray}
\dot\varphi &=& -\nabla.{\bf j}\label{first}\\
{\bf j} &=& - D\nabla\varphi + {\bf V}\varphi \label{second}\\
D(\r) &=& \frac{v(\r)^2\tau(\r)}{d} \label{third}\\
\frac{\bf V(\r)}{D(\r)} &=& -\nabla \ln v(\r) \label{fourth}
\end{eqnarray}
To simplify the form of these equations, we have set $D_t = 0$. They
are the same equations as one would write down to describe a PBP, with a spatially varying $D =
v^2\tau/d$, except for the presence of an extra drift velocity ${\bf
  V}$. This drift velocity is equivalent to an external potential
$\beta U(\r) = \ln v(\r)$, where $\beta \equiv 1/k_BT$.
The active particle behaves as an isothermal PBP in the
presence of this effective potential, which derives solely from
activity \cite{SchnitzerPRE}. Its steady-state probability density
accordingly obeys a Boltzmann-like distribution
\begin{equation}
\varphi_{ss} \propto  \exp[-\beta U] = \frac{1}{v(\r)} \label{steady}
\end{equation} 
Restoring nonzero $D_t$ introduces a factor $(1+D_td/v^2\tau)^{-1}$ in
the r.h.s of~\eqref{fourth}. So long as $v(\r)$ is the only non-constant parameter, this leads to $\varphi_{ss}\propto 1/\sqrt{v^2 \tau+d
  D_t}$. For small
$D_t$ this changes slightly the form of~\eqref{steady}, but
not its qualitative physics.

The effective potential $U$ emerges under conditions where there is no
actual force field acting on the particle. Hence an active origin for
particle diffusion causes deviations from the Boltzmann distribution,
$\varphi_{ss}\propto \exp[-\beta H]$, that cannot be absorbed by any
global rescaling of temperature. (Here $H$ is the actual Hamiltonian,
not incorporating $U$.)  The situation is
somewhat like a PBP in a bath at non-uniform temperature, so that its
(root-mean-square) speed $\bar v$ depends on its
position~\cite{SchnitzerBerg}. Indeed, if one has a box of {\em
  passive} ideal gas particles, in which two halves of the box are
maintained at unequal temperature, then crudely equating the kinetic
particle fluxes from one section to the other across the interface
requires equality of $\rho \bar v$, not of the density $\rho$. In
steady state one therefore recovers $\rho \propto 1/\bar v$ which is
the direct counterpart of \eqref{steady} for a passive, but
non-isothermal, system.

\begin{figure}
\includegraphics{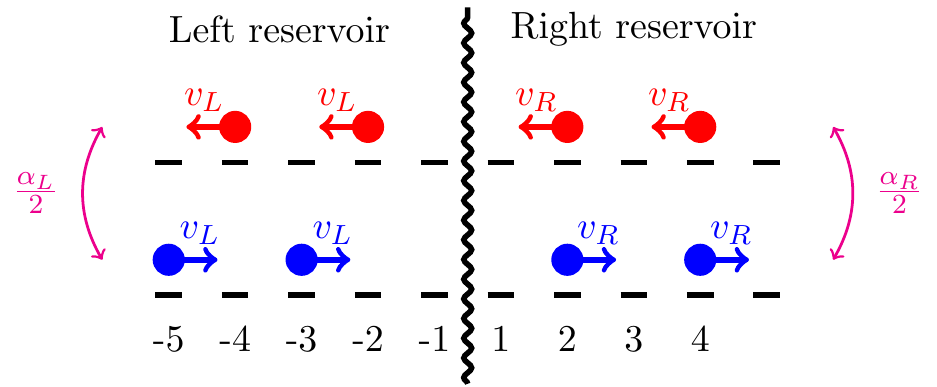}
\caption{Schematic representation of two reservoirs (left and right) containing ideal
  lattice-gases of RTPs in $d=1$ with different speeds $v_L$ and $v_R$. The kinetic flux across
  the boundary is $v_L \rho_{-1}^+-v_R \rho_1^-$, where $\rho_i^{\pm}$
  is the mean number of particles moving to the right or to the left on
  site $i$. (These are shown separately in the figure for the sake of clarity.) In steady-state, where densities are uniform and symmetric
  within each reservoir, the vanishing of the kinetic flux imposes
  $\rho_L v_L=\rho_R v_R$.}
\label{fig:boxes}
\end{figure}

We now give a quantitative version of this argument for an active
ideal 1D lattice gas (Fig.~\ref{fig:boxes}). Here left ($L$) and right ($R$) boxes of RTPs,
having different swim speeds ($v_L,\,v_R$) and tumble rates
($\alpha_L,\,\alpha_R$), are put in contact. (Within each box, left-
and right-moving particles have the same speed and tumble rate; the
notation differs from~\cite{TCPRL}.) Calling $\rho_i^\pm$ the mean
number of particles on site $i$ going to the right ($+$) and to the
left ($-$) we have
\begin{eqnarray}
\label{bulk1}\forall i \neq 1\qquad  \dot \rho_i^+ &=&  v_{L/R} (\rho_{i-1}^+-\rho_i^+)+{\alpha_{L/R}} (-\rho_i^++\rho_i^-)/2\\
\label{bulk2}\forall i \neq -1\qquad   \dot \rho_i^-&=& v_{L/R} (\rho_{i+1}^--\rho_i^-)+{\alpha_{L/R}}(\rho_i^+-\rho_i^-)/2\\
\label{bound1}  \dot \rho_1^+ &=&  v_L \rho_{-1}^+-v_R\rho_1^++{\alpha_R} (-\rho_1^++\rho_1^-)/2\\
\label{bound2}  \dot \rho_{-1}^- &=&  v_R \rho_{1}^--v_L\rho_{-1}^-+{\alpha_L}(\rho_{-1}^+- \rho_{-1}^-)/2
\end{eqnarray}
In steady-state, \eqref{bulk1} and~\eqref{bulk2} impose 
a constant density within each bulk:
\begin{equation}
  \forall i \geq 1 \qquad \rho_{-i}^+=\rho_{-i}^-\equiv \rho_L;\qquad\rho_{i}^-=\rho_{i}^+\equiv\rho_R
\end{equation}
The boundary relations \eqref{bound1}, \eqref{bound2} then require
$v_L \rho_L = v_R \rho_R$, in order to balance the kinetic fluxes from the two reservoirs
across the interface. 
This is a simple instance of the
more general result $\rho \propto 1/v$, and also explains qualitatively why spatial variations of
$\tau$ have no effect on steady states: $\tau$ enters the expression
for the diffusivity, but not directly that for the local flux balance
between our two compartments. Note however that for RTPs with finite
tumble duration $\Delta$,~\eqref{first} and \ref{second}
hold up to the rescaling~$(v,\alpha)\to(v,\alpha)/(1+\alpha\Delta)$~\cite{TCPRL}. The tumbling rate $\alpha$ (and thus $\tau$) then enters
the steady-state distribution since it controls the kinetic flux
through the rescaling of the velocity. This finite-$\Delta$ correction could easily be retained, but is ignored for simplicity from now on.

\section{MANY-BODY PHYSICS}
\label{Sec3}
\subsection{Dynamics of the Collective Density}
Above we addressed a single particle, whose probability density
evolves deterministically via the diffusion-drift equations,
\eqref{first}--\eqref{fourth}. Since these equations also describe a
PBP in an external potential, they are already familiar, and one can
use standard and well-tested procedures \cite{Dean} to derive from
them a stochastic equation of motion for the coarse-grained density
$\rho(\r)$ in a system of many particles. Note that this $\rho$ is not
a probability density (which would evolve deterministically) but a
coarse grained version of the microscopic density operator
$\sum_{i=1}^N\delta(\r-\r_i)$, which obeys a stochastic equation of
motion.  We state the result first for a collection of {\em
  noninteracting} active particles in an environment of spatially
varying motility parameters, $v(\r),\tau(\r)$ \cite{TCPRL}:
\begin{eqnarray}
\dot\rho &=& -\nabla.{\bf J}\label{rhofirst}\\
{\bf J} &=& - D\nabla\rho + {\bf V}\rho + \sqrt{2D\rho}{\bf \Lambda} \label{rhosecond}
\end{eqnarray}
Here $D(\r)$ and ${\bf V}(\r)$ obey Eqs.~\ref{third}, \ref{fourth}, and
${\bf \Lambda}$ is a vector-valued unit white noise. The
multiplicative noise term is to be read in the It\=o sense, which means
that \eqref{rhosecond} is viewed as the small-timestep limit of a
discrete process in which the noise term is evaluated at the start of
each timestep~\cite{Ito}. Another possible choice would be the
Stratonovich convention which requires the noise to be evaluated
mid-way during the timestep. However, switching to this convention
introduces an additional drift velocity into \eqref{fourth} which
is present even for passive particles~\cite{Ito}, making it harder to identify the specific effects of activity. It also makes it harder to generalize to the
interacting case, which is our next task.

\subsection{Density-Dependent Motility Parameters}
\label{sec:vofrho}
Once the It\=o choice is made, \eqref{rhofirst} and \ref{rhosecond} can painlessly be
generalized to the case where the dependence of diffusivity $D$ and
drift velocity ${\bf V}$ on spatial position $\r$ is in fact caused by
a dependence of the motility parameters $v$ and $\tau$ on the density
of particles in that neighborhood \cite{TCPRL}. Of course, this is not the only type of interaction possible: for
instance, hard-core collisions between ABPs are not directly of this
form, but can be partly approximated by it as we shall see in Sec.~\ref{ABPs} below. On
the other hand, bacteria can respond to their local density via a
biochemical pathway (quorum sensing \cite{quorum}), and in some
cases this response is linked directly to their motility
\cite{Hwa1,Hwa2}. Hence it is natural to address the case where the
motility parameters directly depend on the coarse-grained density $\rho$.
Given this choice of interaction, we can write
\begin{eqnarray}
D([\rho],\r) &=& \frac{v([\rho],\r)^2\tau([\rho],\r)}{d} \label{rhothird}\\
\frac{\bf V([\rho],\r)}{D([\rho],\r)} &=& -\nabla \ln v([\rho],\r) \label{rhofourth}
\end{eqnarray}
where the new argument $[\rho]$ denotes an arbitrary functional dependence
on the coarse-grained density field $\rho(\r)$, and we have again set $D_t
= 0$, thereby neglecting the direct Brownian contribution to
translational diffusivity. \eqref{rhofirst} and \eqref{rhosecond} still apply, except that an additional term $\rho \left(\nabla \frac{\delta }{\delta \rho(\r)} D([\rho],\r)\right)$ appears on the right side of \eqref{rhosecond}. In fact this term vanishes in most cases of interest, one exception being the asymmetric lattice model described in Sec.~\ref{lattice} below~\cite{TCPRL}, and we will not discuss it further.

To see whether this description of interacting motile particles is still equivalent to a set of PBPs in thermal
equilibrium, we next consider the Fokker-Planck equation for
the many-body probability ${P}[\rho]$, which reads:
\begin{equation}
  \dot{P}[\rho] = - \int d\r \left(\nabla \frac{\delta}{\delta\rho(\r)}\right) \left[{\bf V} \rho -D\nabla \rho - D\rho\left(\nabla\frac{\delta}{\delta\rho(\r)}\right)\right]{P}[\rho] \label{FPE}
\end{equation}
We may now define an `equilibrium-like' steady
state $P_{\rm eq}[\rho]$ as one in which the probability current vanishes:
\begin{equation}
\label{cond1}
  {\bf \cal J}[P_{\rm eq}]=\left[{\bf V} \rho -D\nabla \rho - D\rho\left(\nabla\frac{\delta}{\delta\rho(\r)}\right)\right]{P}[\rho_{\rm eq}] =0
\end{equation}
Using the ansatz $P_{\rm eq}=\exp[-\beta {\cal F}]$, one sees by
inspection of~\eqref{cond1} that such a flux-free solution exists so
long as the following integrability condition is obeyed:
\begin{equation}
\frac{{\bf V}([\rho],\r)}{D([\rho],\r)}= - \beta \nabla \frac{\delta{\cal F}_{\rm ex}}{\delta\rho}
\end{equation}
which can be rewritten as
\begin{equation}
k_BT \ln v([\rho],\r) \equiv \mu_{\rm ex}([\rho],\r) = \frac{\delta{\cal F}_{\rm ex}}{\delta\rho}
\label{integrable}\end{equation}
This condition requires that the functional defined here as $\mu_{\rm ex}([\rho],\r)$ is itself the derivative of some other functional ${\cal F}_{\rm ex}$. There is no general reason for this to hold.

Nonetheless, if it does hold, our system of interacting active particles is
dynamically equivalent, at large length and time scales, to a PBP
fluid with the free energy functional:
\begin{equation}
{\cal F}[\rho] = k_BT\int \rho (\ln \rho -1) d\r + {\cal F}_{\rm ex}[\rho]
\end{equation}
Here the integral can be viewed as an ideal entropy contribution and
the excess part would, for real PBPs, be caused by some interaction
Hamiltonian. For our active particles, it stems instead from the
density-dependent swim speed $v([\rho],\r)$. Just as in the one-body problem, any density-dependence of the angular
relaxation time $\tau([\rho],\r)$ plays no role in ${\cal F}$. 

Note that the zero-flux condition, \eqref{cond1}, which decides the existence of a steady-state mapping onto a thermal equilibrium system,
cannot be derived without the proper noise terms
in~\eqref{rhosecond}. (These set the prefactor of the second functional
derivative in~\eqref{FPE} to be $D \rho$ and hence lead to the 
condition~\eqref{integrable}.) Unless drift and noise terms are handled on equal terms, for instance when addressing higher order gradient terms of the type discussed in Sec.~\ref{Sec8}, one cannot be sure whether such a mapping still exists or not~\cite{SpeckLowen}.

When it exists, ${\cal F}[\rho]$ specifies not just the most probable configuration of $\rho(\r)$
but also its entire spectrum of steady-state fluctuations, thus taking us far beyond the linear stability analysis sketched
in the introduction.  But, as stated already, one cannot generally
expect \eqref{integrable} to hold true. When it doesn't, one can still formally define some functional ${\cal
  F}=-k_BT\ln P_{\rm stat}[\rho]$, but this no longer governs a flux-free solution
of~\eqref{FPE} and is thus not equivalent to any choice of equilibrium dynamics.
This restates the fact that coarse-graining cannot create a
{\em general} equivalence between active systems, which rely on
microscopically irreversible dynamics, and passive systems, which obey
detailed balance at all scales. 
Suppose however that $\rho$ is slowly
varying in space and that the swim speed $v([\rho],\r)$ depends
isotropically on the values of $\rho$ within some finite range of the point
$\r$. Under these conditions a Taylor expansion gives~\cite{thompson}
\begin{equation}
v([\rho],\r) = v(\rho(\r))+ {\cal O}(\nabla^2\rho) \label{localapprox}
\end{equation}
Suppressing the gradient contribution is equivalent to assuming a {\em
  perfectly local} functional dependence of swim speed on density: $v
= v(\rho)$. As detailed below, it is easy to confirm that ${\cal
  F_{\rm ex}}$, and hence the mapping onto equilibrium, does exist in this case \cite{TCPRL}. This is a limiting
approximation, whose physicality is not
guaranteed. However, much can be gained by assuming the local
approximation, finding the consequences, and then returning to address
gradient effects. The literature has followed this path and
we do so here.\footnote{Should this route not be to the reader's taste, she may skip forward to Sec.~\ref{sec:GradientinF} and then return. That section offers the simplest (phenomenological) treatment of nonlocality, within which all the main results of Sec.~\ref{Sec4} still apply.}

\section{THE LOCAL APPROXIMATION}
\label{Sec4}
\subsection{Without passive Brownian diffusion $D_t=0$}
\label{nodiffusion}
In the local approximation, $v(\r) = v(\rho(\r))$ as just described. The
nonequilibrium excess chemical potential $\mu_{\rm ex} = k_BT \ln v$ then
always obeys the integrability condition, \eqref{integrable} with
${\cal F}_{\rm ex}(\rho) = \int_0^\rho \ln v(s)ds$. The system is equivalent
to interacting PBPs with \begin{eqnarray}
  {\cal F[\rho]} &=& \int f(\rho)d\r \label{localone}\\
  \beta f(\rho) &=& \rho(\ln\rho-1)+\int_0^\rho \ln v(s) ds \label{localtwo}
\end{eqnarray}
Since all steady state statistics depend only on the product $\beta{\cal
  F[\rho]}$, it should by now be clear that we can set $\beta = 1$ without
loss of generality, and we do this silently from now on.
The chemical potential may then be written as a sum of ideal and
excess parts
\begin{equation}
\mu = \frac{\delta {\cal F}}{\delta{\rho(\r)}}= \mu_{\rm id} +\mu_{\rm ex} = \ln \rho(\r) + \ln v(\rho(\r)) \label{mulocal}
\end{equation}
It should be noted that, within the premise of interactions whose sole
effect is to make $v$ and $\tau$ depend on density,
Eqs.~\ref{localone}-\ref{mulocal} involve no approximation beyond
the locality of $v(\rho)$ and the validity of the coarse-graining
leading to the mesoscopic equations \ref{rhofirst}-\ref{FPE}.

Nonetheless, we first proceed within the context of a (Landau-like)
mean-field theory in which spatial fluctuations are ignored,
performing a global minimization of the function
$F=\sum_iV_if(\rho_i)$ at constant volume $\sum_iV_i$ and constant
particle number $\sum_iV_i\rho_i$: phase separation then arises
whenever more than one $V_i$ is nonzero. An important caveat is that
this makes sense only if the interfacial tension between phases is
finite, since this tension alone prevents the macroscopic phases from
fragmenting into uncountably many small domains at the (nominally)
coexisting densities. This means that the global minimization, while
not making direct use of gradient terms, tacitly assumes that these
terms exist and do not violate \eqref{integrable}. We proceed on that
basis, but revisit this issue in Sec.~\ref{Sec8}.

\begin{figure}
\begin{center}
  \begin{minipage}{.49\textwidth}
    \includegraphics[width=5cm]{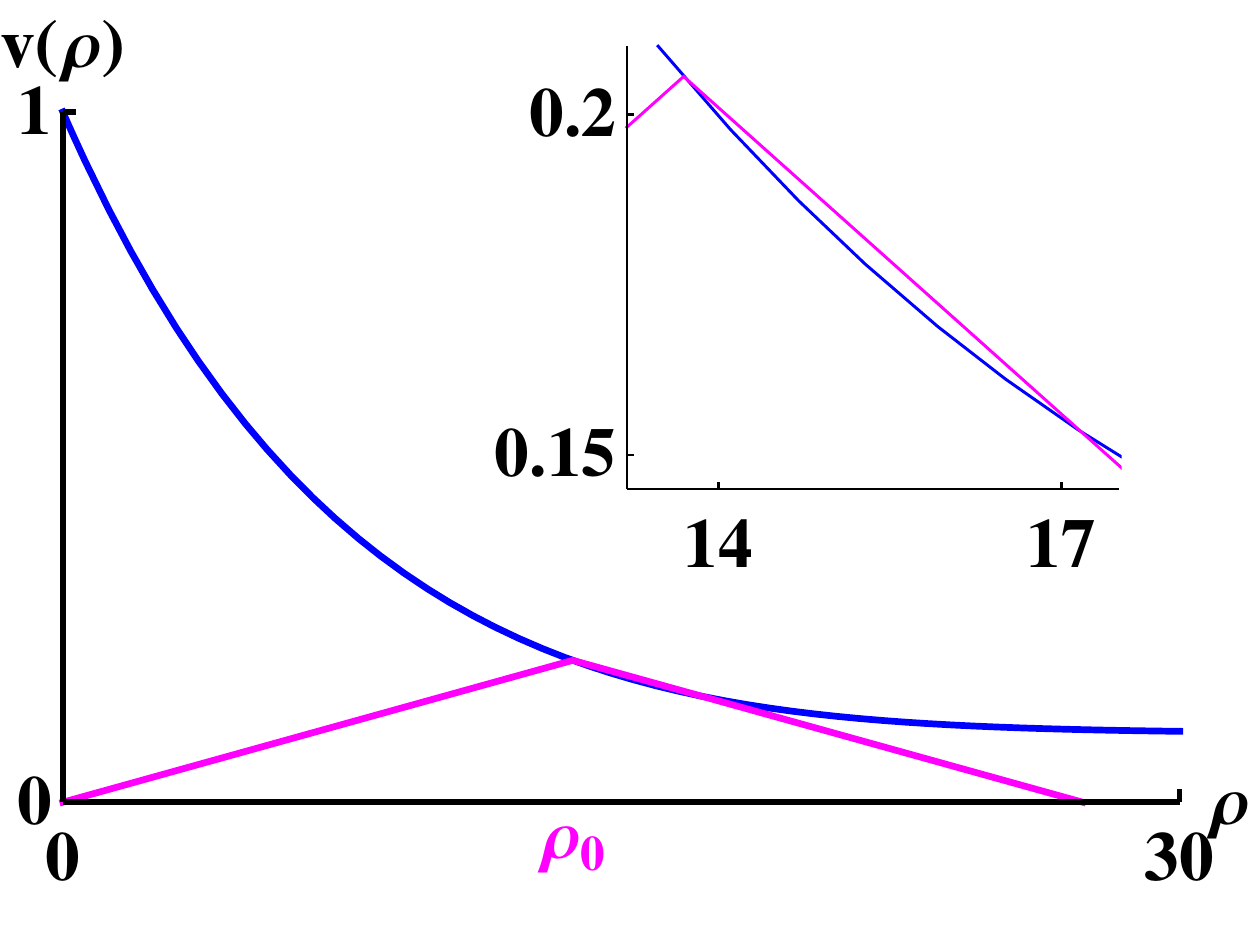}
  \end{minipage}
  \begin{minipage}{.49\textwidth}
    \includegraphics[width=5cm]{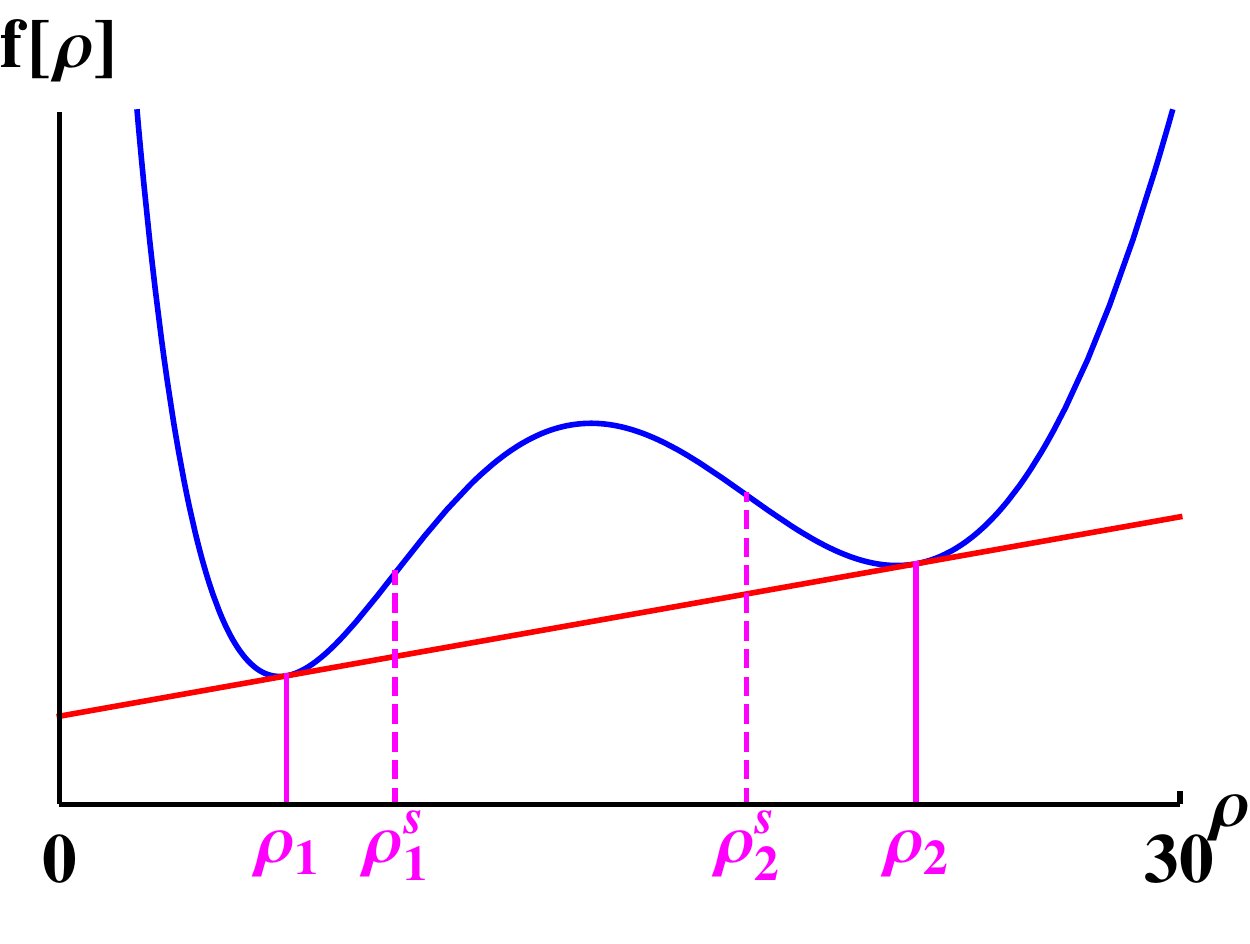}
  \end{minipage}

\caption{Construction of the effective free energy density $f(\rho)$
  in the mapping from active particles with strictly local motility
  interactions onto a fluid of interacting Brownian particles. If
  $v(\rho)$ decreases rapidly enough (left) the resulting $f(\rho)$
  has a negative curvature (spinodal) region (right) with the global
  equilibrium state comprising a coexistence of the binodal densities
  $\rho_1,\rho_2$. The condition for instability ($f''<0$) translates
  into the geometric construction shown on $v(\rho)$: draw a line from
  the origin to any point on the curve and reflect this line in the
  vertical axis. If the slope of $v(\rho)$ is less than the reflected
  line, the system is unstable. The figures correspond to $v(\rho)$
  given by \eqref{eqn:vofrhoexpo} with $D_t\simeq 1.4\ 10^{-3}$,
  $v_0=1$, $v_1=0.1$, $\varphi=4$, $\tau=2$, $d=2$ and $2
  \rho_0=\rho_1^s+\rho_2^s$.}

 \label{CTC}
\end{center}
\end{figure}

For two-phase coexistence, the global minimization proceeds just as in equilibrium: one first searches for concavities in
the function $f(\rho)$, and where these exist performs a
common-tangent construction. This construction finds unique coexisting
densities $\rho_1$ and $\rho_2$, such that the chemical potential $\mu
= df/d\rho$ in the two phases are equal, and such that their
thermodynamic pressures $p = \mu\rho-f$ are also equal. The first of
these equalities requires equal slopes for the tangents to $f(\rho)$
at the coexisting densities, whereas the second requires equal
intercepts: thus a single tangent connects both coexistence densities
(Fig. \ref{CTC}).  The condition for concavity, $f''(\rho)<0$ (where
primes now denote $\rho$ derivatives), implies $v'(\rho) < -v/\rho$,
which is exactly the condition for linear instability found in~\eqref{eqn:InstabCriterion}. The set of such negative-curvature points defines the
spinodal region, a familiar concept in mean-field thermodynamics
\cite{spinodal}. The common-tangent construction also encloses zones
where $f$ has positive curvature; here phase separation occurs by
nucleation and growth \cite{Bray}.

This approach thus shows how self-propelled particles with no attractions but
a decreasing $v(\rho)$ can be mapped, at a (Landau-like) mean-field level of global minimization, to a system
of attractive PBPs undergoing equilibrium liquid-gas phase separation.

\subsection{The effect of finite thermal diffusivity: $D_t\neq 0$}
\label{translationaldiffusion}
We have just outlined the theory of motility-induced phase separation
(MIPS) as first presented in \cite{TCPRL} for RTPs, and extended later
to include continuous angular diffusion in \cite{EPL2}, thereby embracing ABPs also. For simplicity we neglected
the Brownian translational contribution
$D_t$; restoring this (as discussed after \eqref{steady} above) gives the modified integrability condition
\begin{equation}
\frac{\tau v \nabla v}{v^2 \tau+d D_t} = \nabla \frac{\delta }{\delta \rho} {\cal F}_{\rm ex}
\end{equation}
For a strictly local
$v(\rho(\r))$,
a sufficient condition for ${\cal F}_{\rm ex}$ still to exist is that {\em only} the speed (not $\tau$ or $D_t$) depends on the density.  One then find an excess free energy
\begin{equation}
  {\cal F}_{\rm ex}=\int f_{\rm ex}(\rho(\r)) d\r\qquad\mbox{with}\qquad f_{\rm ex}(\rho)=\int^\rho \frac{1}{2} \ln[v^2(s) \tau+d D_t]  d s
\end{equation}
Again, the non-convexity of the free energy density $f(\rho)=\rho
(\ln \rho-1)+f_{\rm ex}(\rho)$ signals the possibility of MIPS; 
the spinodal region corresponds to
\begin{equation}
\label{eqn:instabDt}
  f''(\rho)<0 \qquad\Leftrightarrow\qquad  {v^2 \tau} \left( 1+\rho\frac{v'}{v}\right) <-{d D_t}
\end{equation}
For $D_t=0$, one recovers~\eqref{eqn:InstabCriterion}, whereas a
finite $D_t$ makes the system more stable by tending to smooth out
density fluctuations. In particular, when $v\to 0$, the modified
condition~\eqref{eqn:instabDt} is never fulfilled if $D_t$ is
finite. For any given $v(\rho)$, there is thus a minimal ratio ${v^2
  \tau}/{d D_t}$ of active to thermal diffusivities below which MIPS
never occurs.

As an example, let us consider a propulsion speed decaying exponentially from
a value $v_0$ to a smaller one $v_1$, on some characteristic density
scale $\varphi$:
\begin{equation}
  \label{eqn:vofrhoexpo}
  v(\rho)^2=v_0^2+(v_1^2-v_0^2) (1-e^{-\rho/\varphi})
\end{equation}
With this choice, $f_{\rm ex}(\rho)={\rho} \ln[v_1^2 \tau +d D_t]/2
-{\varphi}\left[{\rm Li}_2(-A)-{\rm Li}_2(-A e^{-\rho/\varphi})
\right]/2$ where ${\rm Li}_2(z)$ is the polylogarithm function of
order 2 and $ A=({v_0^2 \tau -v_1^2 \tau})/({v_1^2\tau+d D_t})$.  The
term in $f_{\rm ex}$ that is linear in $\rho$ plays no role in
stability or phase equilibria. The spinodals ($f''(\rho)=0$) are roots
of $\lambda \exp\lambda = -2e^2/A$, where $\lambda \equiv
2-\rho/\varphi$, which exist only when $A\ge 2e^3/3$. 
For a given $v(\rho)$, this sets a
maximal value for $D_t$
\begin{equation}
  D_t^c=\frac{v_0^2 \tau-v_1^2\tau(1+2 e^3)}{2 d e^3}
\end{equation}
at which MIPS ends, presumably at a critical point. Conversely, for
given $v_1$ and $D_t$, there is a minimum bare swim speed $v_0^c$ for MIPS, obeying
\begin{equation}
  v_0^c=\sqrt{v_1^2 (1+2e^{3})+2e^3 \frac{D_t d}{\tau}}
\end{equation}

\begin{figure}
  \begin{minipage}{.49\textwidth}
    \includegraphics[width=5cm]{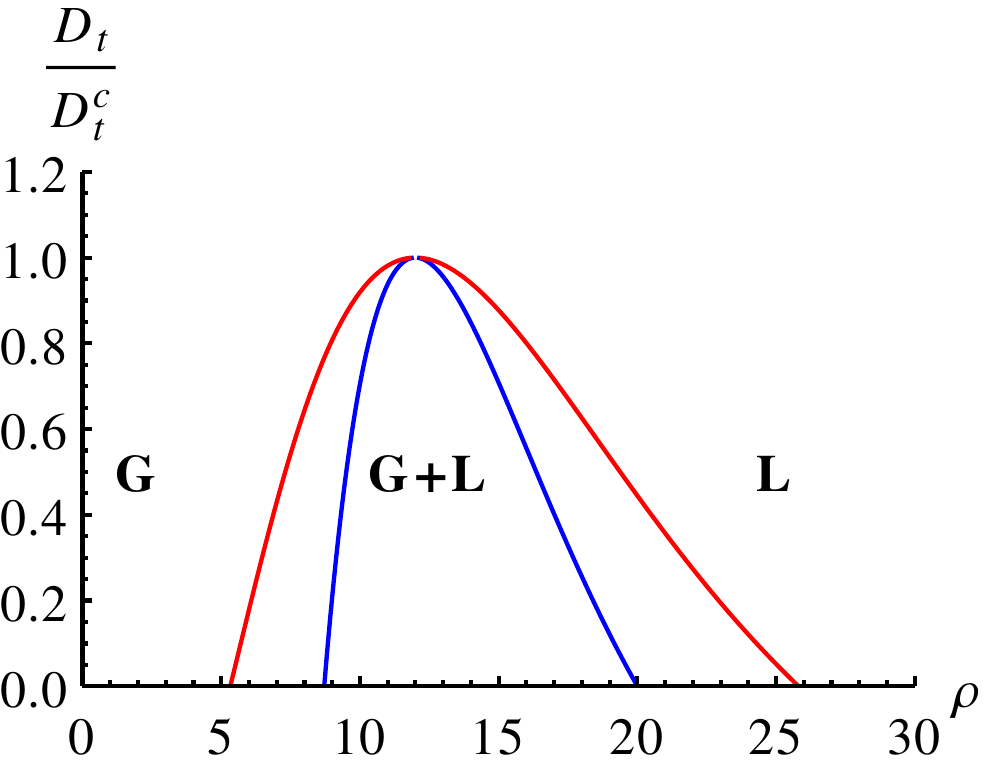}
  \end{minipage}
  \begin{minipage}{.49\textwidth}
    \includegraphics[width=5cm]{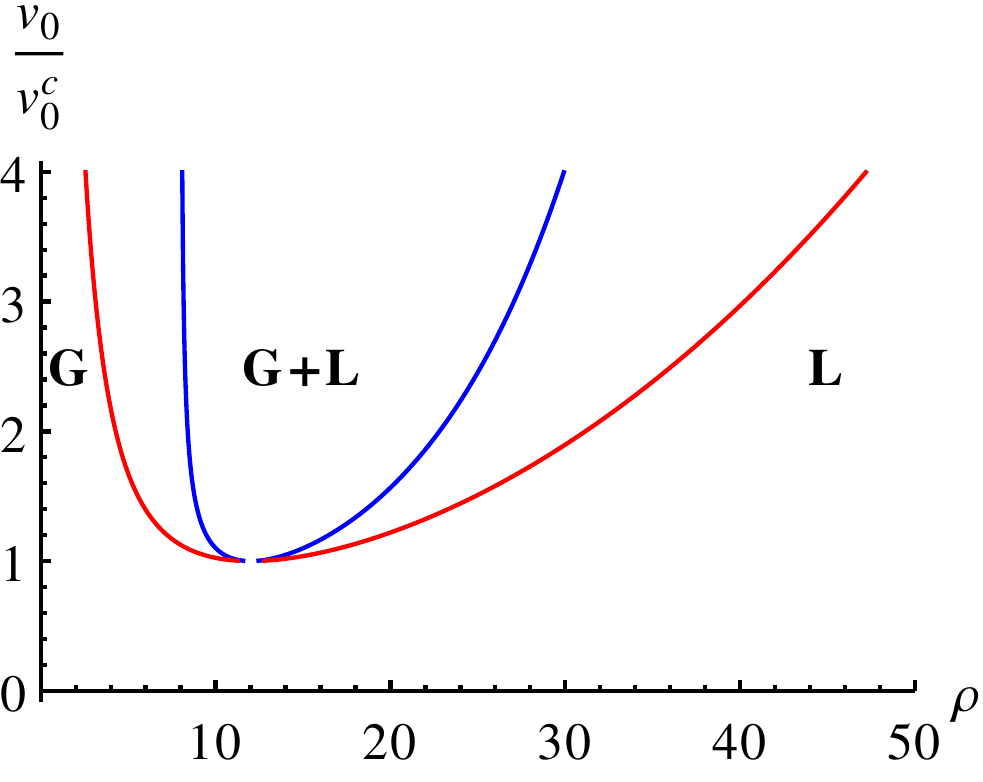}
  \end{minipage}
  \caption{Liquid-gas phase diagrams of MIPS for $v(\rho)$ as defined
    in~\eqref{eqn:vofrhoexpo} with $v_0=1$, $v_1=0.1$, $\varphi=4$,
    $d=2$, $\tau=1$ (left) and $D_t=0.25$, $v_1=0.25$ , $\varphi=4$,
    $d=2$, $\tau=1$ (right). Red and blue lines correspond to binodal
    and spinodal curves.}
\label{fig:SL}
\end{figure}
\begin{figure}
  \begin{minipage}{.49\textwidth}
    \includegraphics[width=5cm]{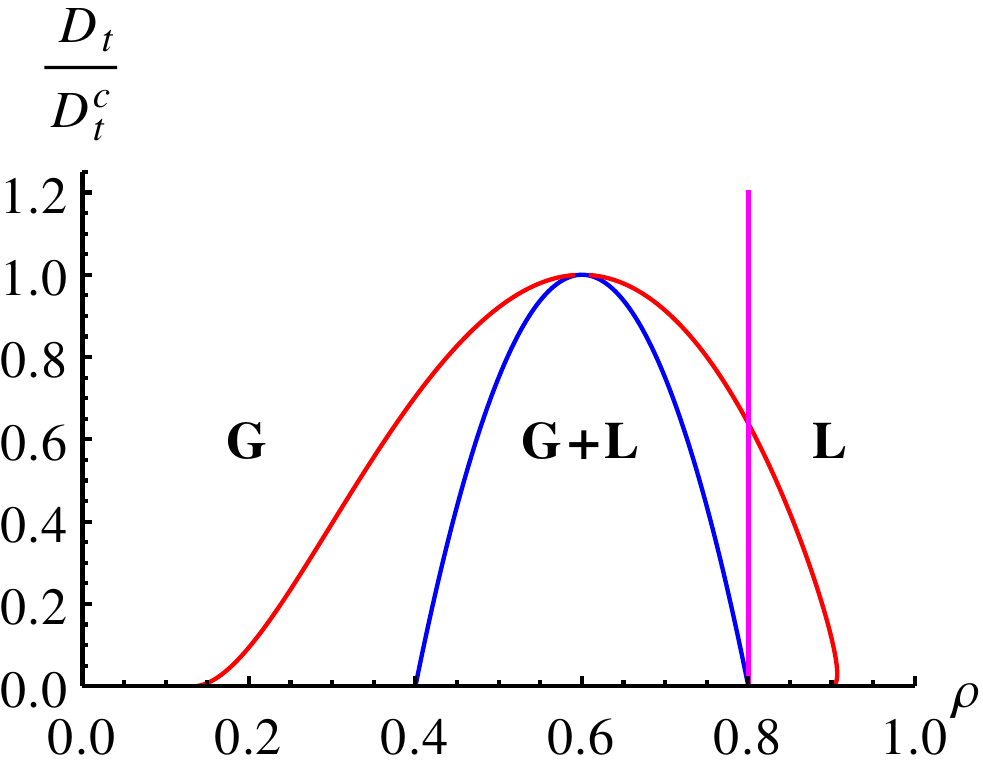}
  \end{minipage}
  \begin{minipage}{.49\textwidth}
    \includegraphics[width=5cm]{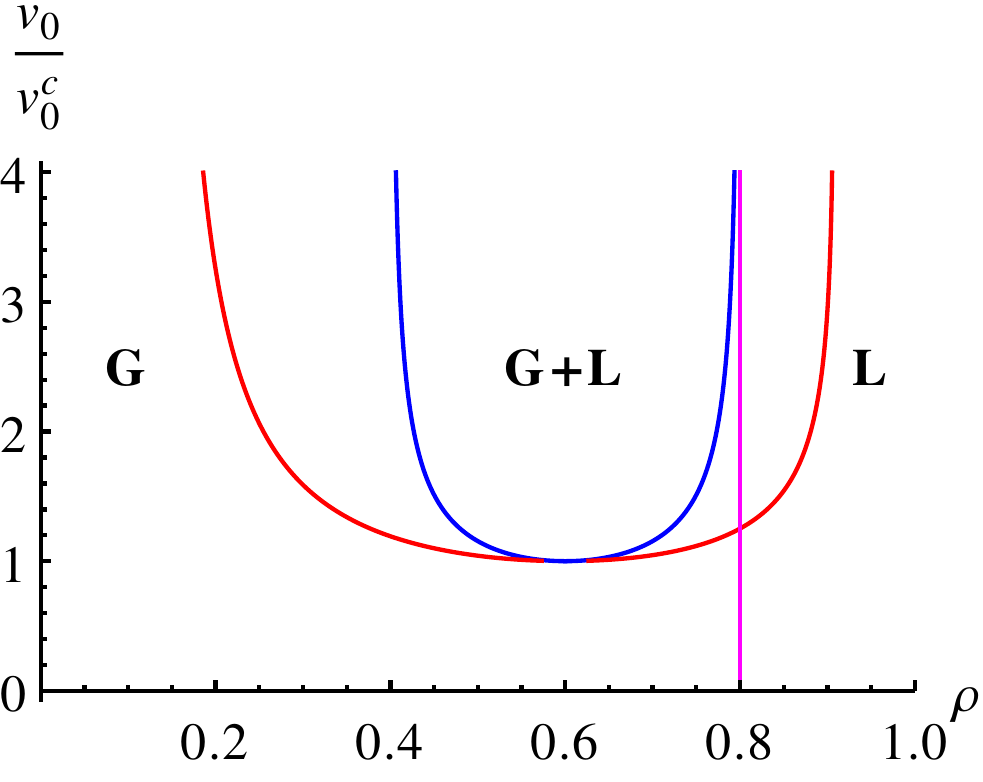}
  \end{minipage}
  \caption{Phase diagrams of MIPS for $v(\rho)$ as defined
    in~\eqref{eqn:vofrholinear} with $v_0=5$ (left), $D_t=0.5$ (right), $d=2$, $\tau=1$ and $\rho^* = 0.8$ (vertical
    line). Red and blue lines are binodal and spinodal curves.}
\label{fig:vofrholinear}
\end{figure}

The spinodal and coexistence lines for $v(\rho)$ obeying \eqref{eqn:vofrhoexpo} are
shown in Fig.~\ref{fig:SL}, while Fig.~\ref{fig:vofrholinear} gives those for an even simpler form of $v(\rho)$, which features below (see \eqref{linear}):
\begin{equation}
\label{eqn:vofrholinear}
  v(\rho)=v_0(1-\rho/\rho^*)
\end{equation}
Again the excess free energy can be computed exactly\footnote{Specifically, 
$
f_{\rm ex}=\frac{\rho}2(\ln(v_0^2 \tau)-2) +\frac{\rho-\rho^*}{2}\ln\Big[\epsilon+\Big(1-\frac{\rho}{\rho^*}\Big)^2\Big]- \rho^* \sqrt{\epsilon} \arctan\left(\frac{1-\frac{\rho}{\rho^*}}{\sqrt{\epsilon}}\right)
$.} 
as can the spinodals:
\begin{equation}
\label{linearspinodal}
\rho_s^\pm={\rho^*} \left(3\pm \sqrt{1-8\epsilon}\right)/4
\end{equation}
where $\epsilon={d D_t}/{v_0^2 \tau}$ is a ratio of
passive and active diffusivities.
MIPS thus exists only when $\epsilon<1/8$, i.e.
\begin{equation}
  D_t <  D_t^c\equiv \frac{v_0^2 \tau}{8 d};\qquad\mbox{or}\qquad v_0\geq v_0^c\equiv\sqrt{ \frac{8 d D_t}{\tau}}
\label{criticallinear}  
\end{equation}
Note that \eqref{eqn:vofrholinear} for
$v(\rho)$, and also the resulting phase diagrams, are only meaningful for $\rho\leq \rho^*$, beyond which the system reverts to a passive state (vertical lines in Fig.~\ref{fig:vofrholinear}). 

The study of the spinodals and the limit of existence of MIPS below a
critical velocity can also be derived through a linear stability
analysis of continuum mean-field
equations~\cite{BialkeLowen,HenkesSoft}. Predicting the binodals,
however, requires the derivation of the effective free energy.

\section{NUMERICAL EVIDENCE FOR MIPS}
\label{Sec5}
\subsection{Phase Separation in Run-and-Tumble Particles}
\label{lattice}
Clear numerical evidence for MIPS within the region where $v'<-v/\rho$
(\eqref{eqn:InstabCriterion}) was presented in \cite{TCPRL} for
simulations of RTPs in $d=1$ dimension. In these simulations, a fixed
coarse-graining length is used to define the density $\rho$ upon which
$v$ then depends locally.  The coexisting densities were compatible,
within numerical error, with those set by the common-tangent
construction.

It is known that for the equivalent system of PBPs with attractions,
phase separation cannot proceed to completion in $d=1$, because domain
walls have a finite energy cost $E$. The equilibrium state thus has a
domain-wall density $\propto\exp[-\beta E]$. While this precludes
long-range order, for large $E$ the `phase-separated' state
(containing sparse domain walls between patches of $\rho_1$ and
$\rho_2$) is unambiguously different in appearance from the
`single-phase' regime in which there are only small density
fluctuations about an average value.  The 1D simulations of \cite{TCPRL}
observed phase separation in this restricted sense; domain walls were
formed and slowly became more dilute through a coalescence process.

Microscopic simulations of RTPs with crowding interactions were
subsequently presented in $d=1$ and $d=2$ using a fast lattice-based
discretization~\cite{thompson}.  In $d=1$
the hopping rate of right-going ($+$) and left-going ($-$) particles on site $i$ can be
chosen as 
\begin{equation}
\label{RTPOL}
  v^{\pm}(i,[\rho])=v_0\left(1-\frac{\tilde \rho_i}{\rho_{M}}\right);\qquad\mbox{with}\qquad \tilde\rho_i=\sum_j K^{\pm}_{ij} \rho_j
\end{equation}
Here $\rho_j$ is the number of particles on site $j$, $\rho_M$
controls the maximal number of particles on each site, and $K_{ij}$ is
a kernel describing how particles on site $i$ interact with those on
site $j$.  When $K_{ij}$ is a smooth symmetric kernel, particles are equally sensitive to the
particles in front of them or behind them. (This seems appropriate for chemically mediated interactions.) Conversely, for asymmetric
kernels the particles are more sensitive to those in front of them;
the limiting case $K^{\pm}_{ij}=\delta_{i\pm 1,j}$ corresponds to a partial
exclusion process, which is a good lattice-based model of steric
crowding~\cite{Sandow1994,TailleurJPA}.

A fluctuating hydrodynamic description akin
to Eqs.~\ref{rhofirst}--\ref{rhosecond} can be derived to account for
the large scale behavior of this system~\cite{thompson}. This shows a
mapping to an equilibrium system of passive particles with attractive
interactions. Indeed the system admits a free energy functional ${\cal
  F}$ which predicts MIPS exactly as presented in
Sec.~\ref{Sec4}~\cite{thompson}. Homogeneous profiles are predicted to
be linearly unstable only at rather high densities ($\rho>\rho_M/2$)
but nucleation can occur at a much lower ones~\cite{thompson}.  In
$d=1$, MIPS again creates alternating sequences of high- and
low-density domains, separated by domain walls. (The limiting case
$\rho_M=1$ with $K_{ij}^\pm=\delta_{i\pm 1,j}$ has been studied in
detail in~\cite{golestanianRTP}.)

\eqref{RTPOL} can easily be generalized to higher dimensions and we
now turn to the 2D case. For symmetric kernels $K_{ij}$, in the limit
of large $\rho_M$, the mean-field free energy analysis predicts
quantitatively both the occurrence of complete phase-separation and
the values of the coexisting densities~\cite{thompson}. When $\rho_M$
is finite, MIPS still occurs but (as is common for partial exclusion
models \cite{TailleurJPA}) the coexisting densities are not those
predicted by mean-field theory.  The case of an asymmetric kernel,
which might be thought to mimic steric exclusion on a lattice more
closely, has a richer phenomenology. MIPS occurs, but the coexisting
densities are incorrectly predicted by the common-tangent construction
even when mean-field theory might be expected to hold
($\rho_M\to\infty$). For instance, the gas density in the phase
coexistence region, given by the lower binodal,  goes to
zero as $v_0\to \infty$~\cite{thompson} while the theory of
Sec.~\ref{Sec4} predicts a non-vanishing saturation vapor density in this
limit.

\begin{figure}
  \begin{center}
    \begin{minipage}{2in}
      \includegraphics[width=2in]{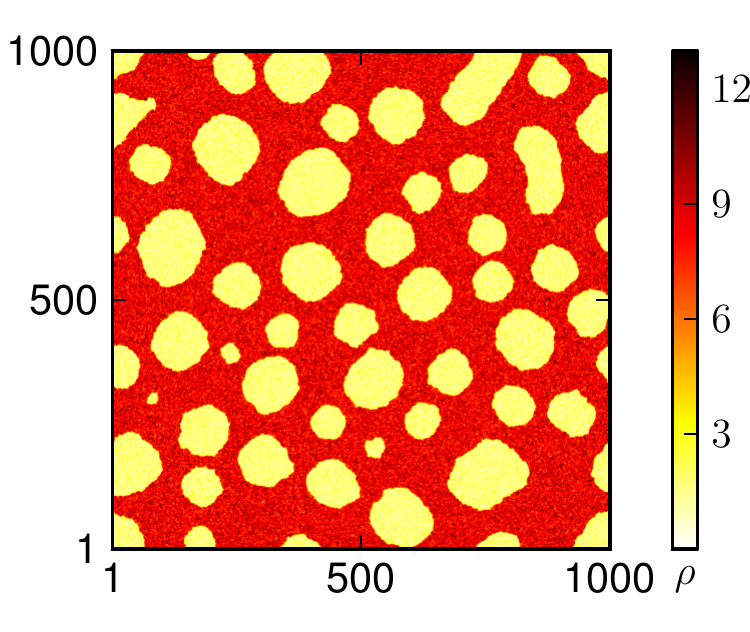}
    \end{minipage}
    \hspace{1cm}
   \begin{minipage}{5cm}
      \includegraphics[width=5cm]{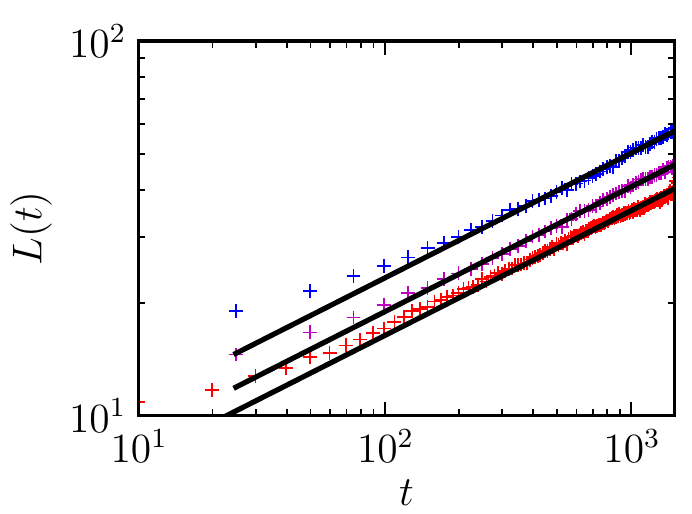}
   \end{minipage}
  \end{center}
  \caption{(Left) A 2D run-and-tumble system undergoing
    MIPS. Simulated via a lattice model ($1000\times 1000$ sites) as
    detailed in \cite{thompson}, with local density (particles per
    site) color-coded on scale at right. (Right) Log-log plot of
    domain scale $L(t)$ for droplet coarsening within the spinodal
    region. Solid lines have slope 1/3. {Results for three different
    kernels relating $v$ to the density are shown for $200\times 200$
    site lattices (adapted from~\cite{thompson}, courtesy of A. Thompson).}}\label{fig:ThomsonFig}
\end{figure}

Furthermore, for the asymmetric kernel $K_{ij}^\pm=\delta_{i\pm 1,j}$,
MIPS is only seen for large enough values of $\ell=v_0\tau$; when this
is less than about four times the repulsion radius (four lattice
sites), homogeneous profiles are instead stable. This is related to,
but distinct from, the minimum-speed requirement for off-lattice MIPS
found above in Sec.~\ref{Sec4}.  In the lattice models, the coarse-grained
theory of MIPS fails to capture the minimal run-length, not because
$D_t$ is neglected, but because of the short-scale breakdown of the
diffusive approximation itself (\eqref{rhofirst} and \ref{rhosecond}).
By formulating a mean-field theory directly at
the microscopic lattice level~\cite{thompson}, a minimum speed
criterion is recovered. This breakdown of MIPS was seen for general anisotropic kernels $K$, but not for rotationally symmetric ones. 
Since these kernels differ only at the level of gradient terms so far
neglected, this points to a more important role for such terms than in
equilibrium problems (see Section~\ref{Sec8}).

In all 2D systems where MIPS is seen, an initially homogeneous profile
in the spinodal region gives way to a droplet domain morphology whose
length scale $L(t)$ coarsens with a power law in time: $L(t)\sim
t^{1/3}$, see Fig.\ref{fig:ThomsonFig}. This is the classically
expected result for diffusive coarsening, without coupling to a
momentum-conserving fluid, in a passively phase-separating
system~\cite{Bray}. The exponent can be explained by considering the
diffusive flux down chemical potential gradients set by the Laplace
pressure differences across curved interfaces of radius $L$ and fixed
interfacial tension. Since this tension vanishes within the local
approximation, we defer further discussion to Section \ref{Sec8}.

\subsection{Phase Separation in Active Brownian Particles}
\label{ABPs}
Convincing evidence for MIPS is also seen in numerical
studies on ABPs for a variety of hard-sphere
potentials~\cite{RednerPRL,BialkeLowen,StenhammarPRL,StenhammarSoft,Gompper},
as well as for soft spheres~\cite{Fily,HenkesSoft}. Apart
from the difference in rotational diffusion dynamics, which is
inessential \cite{EPL2}, these simulations crucially differ from those
on RTPs in which a density-dependent swim-speed is directly encoded
into the dynamics. Simulations on ABPs generally instead address
hard-core swimming particles, whose $v([\rho],\r)$ is not encoded {\em
  a priori} into the equations of motion (an exception being~\cite{Ludo}). Instead,
collisions can be expected to slow down the particles at
high density. One can then define an \textit{emergent} $v$ as the average of
the true particle velocity projected along the propulsion
direction. Monitoring this within bulk systems of uniform density
$\rho$, one finds $v(\rho)$ decreases almost linearly with density
\cite{Fily,RednerPRL,StenhammarPRL,StenhammarSoft}:
\begin{equation}
v(\rho) =  v_0[1-\rho/\rho^*] \label{linear}
\end{equation}
where $v_0$ is the dilute swim speed and $\rho^*$ the extrapolated
point at which $v$ vanishes altogether. The latter is barely
distinguishable from the close packing threshold in both $d=2$
and $d=3$ \cite{Fily,StenhammarPRL,StenhammarSoft,Gompper}, although this threshold is influenced by activity \cite{BerthierPack,Henkes}, and also by slight softness of the
particles in the simulations. Eq.~\ref{linear} is not only confirmed
to high accuracy in simulations but also predicted by various types of
kinetic theory
\cite{Fily,BialkeLowen,HenkesSoft,StenhammarPRL}. Indeed its linear
form is easily deduced from a binary collision picture in which a
particle stalls for a fixed time interval during each collision.

In contrast to the observed linear behavior of $v(\rho)$, most
simulations of spherical ABPs allow no dependence whatsoever
of $\tau$, the rotational relaxation time, on density $\rho$. This is
because collisions, caused by pairwise central forces between
particles, cannot exert torques and so cannot rotate the swimming
direction. Hence $\tau^{-1} = (d-1)D_r$ at all densities.  The ABP
simulations then confirm that the particle diffusion constant obeys
$D(\rho) = v(\rho)^2\tau/d+D_t$ as expected, with $v(\rho)$ obeying
Eq.~\ref{linear} \cite{Fily,RednerPRL,BialkeLowen,StenhammarPRL,StenhammarSoft}.

To test the relevance of the MIPS theory of Sec.~\ref{nodiffusion}
for an ABP system, one must note that the density cannot exceed
$\rho^*$, the close packed value. This stems from hard sphere
compressibility constraints that are not encoded in $v(\rho$): at high density $v$ is almost zero, and ${\cal F}$ for \eqref{localtwo} reverts to that of a highly compressible, passive
ideal gas. Thus the mean-field free
energy density $f(\rho) = \rho(\ln\rho-1) + \int_0^\rho\ln v(s)ds$ has
to be supplemented by an additional constraint that imposes
$\rho\le\rho^*$ \cite{StenhammarPRL}.
Once this is done, both steady-states and coarsening dynamics seen in
ABP simulations are in qualitative agreement with the theory of
MIPS~\cite{StenhammarPRL,StenhammarSoft}. However, some important
physics is missing, in that the phase equilibrium predicted from
\eqref{linear} is independent of the P\'eclet number, here defined as
\begin{equation}
\hbox{\rm Pe}  = \frac {3 v_0\tau}{\sigma} \label{Peclet}
\end{equation}
where $\sigma$ is the particle diameter.  No MIPS is seen numerically
for $\Pe<\Pe_c$, with $\Pe_c \simeq 55$ in $d=2$ and $\simeq 125$ in
$d=3$; instead the predicted two-phase region closes off at
$\Pe=\Pe_c$, probably ending in a critical point
\cite{Fily,RednerPRL,StenhammarPRL,StenhammarSoft,RednerPRE}.  On the
contrary, the MIPS theory of Sec.~\ref{nodiffusion} always predicts a
spinodal decomposition for $\rho>\rho^*/2$.

Since $\tau$ arises by thermal diffusion (in ABPs), the P\'eclet
number not only governs the ratio of the persistence length
$\ell = v_0\tau$ of a dilute swimmer to its diameter $\sigma$, but
also is of order $v_0/v_0^c$ where $v_0^c$ is the speed threshold for MIPS set by translational diffusion (see
\eqref{criticallinear}). Thus in experimental ABPs, Pe 
controls two distinct mechanisms for loss of MIPS: translational
thermal diffusion (described in Sec.~\ref{translationaldiffusion}) and
a small persistence-length to diameter ratio (discussed, for lattice
models, in Sec.~\ref{lattice}). In simulations, however, these can be
distinguished by artificially setting $D_t = 0$. The theory of MIPS
laid out in Sec.~\ref{nodiffusion} ignores both those effects; it is thus
not suprising that this limiting theory thus does not capture the
observed disappearance of MIPS in ABPs as Pe is reduced.

The existence of a critical P\'eclet number $\Pe_c$ is consistent with various approaches based on the kinetic theory of gases, suitably
adapted to ABPs~\cite{ButtinoniBechinger, 
Fily, RednerPRL, BialkeLowen, SpeckLowen,HenkesSoft, StenhammarPRL, Henkes,RednerPRE}. These approaches
yield continuum equations whose linear stability analysis can be used
to locate
spinodals~\cite{Fily,BialkeLowen,HenkesSoft}. The results resemble the phase diagrams predicted by the 
theory of MIPS presented in Sec.~\ref{translationaldiffusion}, which
allowed for thermal translational diffusion ($D_t \neq 0$).
However, whether a finite $D_t$ is indeed what suppresses phase separation at small $\Pe$ is questionable.
Since $\Pe\sim v/v_0^c$, MIPS should then persist down to Pe of order unity
(see Fig.\ref{fig:vofrholinear}), thereby under-predicting the
reported $\Pe_c$ for MIPS by a factor of 50 or more.
This is strong evidence that nonzero $D_t$ is not solely responsible
for the loss of MIPS at low $\Pe$ in ABPs. Reinforcing this is the
observation that simulations in which $D_t$ is set to zero
\cite{Fily,HenkesSoft} give very similar results to those in which
$D_t$ takes the thermal value set by matching the Brownian mechanism
for angular rotation \cite{RednerPRL,StenhammarPRL,StenhammarSoft}.

One interesting alternative avenue is to use the rate of arrival and
departure of ABPs at the surface of a dense domain to compute
kinetically the vapor density $\rho_v$~\cite{RednerPRL,RednerPRE}. This approach predicts the
existence of a critical $\Pe$. The arrival rate of particles (per unit area) is of
order $\rho_vv_0$ while the departure rate is $\kappa D_r/\sigma$,
which involves $D_r$ because arriving particles must rotate through a
finite angle before they can leave. Here $\kappa$ is a dimensionless
(fitting) factor that allows for the fact that particles tend to leave
in bursts rather than individually. Equating rates gives an expression
for the fraction $f_c(\rho,\Pe)$ of particles in clusters; contours of
this function vary as $\Pe^{-1}$, and the limit $f_c=0^+$ closely
tracks the simulated binodal up to $\Pe\simeq 100$. However, this
means $f_c(\rho,\Pe) \to 1$ for all $\rho$ as $\Pe\to\infty$, so there
is no connection in that limit with the phase-separation theory of
MIPS that assumes a smooth $v(\rho)$. A partial reconciliation between
the two approaches could be to follow~\cite{HenkesSoft} and make
$\rho^*$ in \eqref{linear} an explicit function of $\Pe$.

In summary, the `thermodynamic' approach to MIPS in ABPs (based on
assuming a smooth $v(\rho)$) can be qualitatively improved by allowing
for finite rotational P\'eclet number, as in
Sec.~\ref{translationaldiffusion}. It then yields spinodals in
agreement with kinetic theory arguments~\cite{BialkeLowen,HenkesSoft}.
This is however not the main correction needed to account for ABP
simulations, which instead stems from the discreteness of the
collisional dynamics{, as for the lattice model of Sec.~\ref{lattice}}. This
can be partially understood using a kinetic approach to cluster
growth~\cite{RednerPRL,RednerPRE} but, so far, this has not been
married to the thermodynamic picture, and we do not yet have a
complete theory for the critical P\'eclet number $\Pe_c$, below which
MIPS disappears.

The region close to the critical point, and any universal exponents
associated with it, also remains to be explored (theoretically or
numerically), in both $d=2$ and $d=3$. Continuum theories (discussed
in Section \ref{continuum} below) could be useful here, as the direct
simulations of active particles already require extremely large
systems (up to 40 million particles in \cite{StenhammarSoft}) to
resolve even the non-critical aspects of MIPS that prevail at high
P\'eclet number.

Such aspects include the coarsening law for the domain size $L(t)$
after a quench into the phase-separated regime. {In two dimensions this
shows an exponent $L(t)\sim t^\alpha$ with $0.25 \le \alpha \le 0.28$
\cite{RednerPRL,StenhammarPRL,StenhammarSoft,RednerPRE}. This is
somewhat below the value of $1/3$ expected for passive coarsening (and
which was also reported in \cite{thompson} for RTPs in 2D and
in~\cite{StenhammarSoft} for ABPs in 3D).} However it is not yet clear
that the difference is numerically significant; it may instead reflect
a slow transient approach to an asymptotic $1/3$ power
\cite{Wittkowski} (see Section \ref{continuum} below). {All in
  all, the numerical results on
  ABPs~\cite{RednerPRL,StenhammarPRL,StenhammarSoft,RednerPRE} and
  RTPs~\cite{thompson} for the coarsening exponents call for
  complementary studies on larger system sizes. Furthermore,} even if
the coarsening behavior of ABPs broadly resembles that of a passive
system, this is not true when one looks in more detail. For instance
in $d=2$ \cite{StenhammarPRL} and also in $d=3$ \cite{Gompper}, one sees `lava-lamp' type dynamics in
which even fast local fluctuations within a well-separated domain
manifestly break time reversal symmetry. This contrast with the
passive case where irreversible dynamics is visible only at scales
above $L(t)$.

\section{EXPERIMENTAL IMPLICATIONS}
\label{Sec7}
\subsection{Experiments on Bacteria}
In microbiological studies, formation of dense clusters from a uniform initial population of motile bacterial cells is often encountered (and usually called `aggregation' rather than phase separation). So far though, a quantitative link between this behavior and MIPS has not been established.

The down-regulation of swimming activity at high density is fundamental to the formation of biofilms \cite{biofilms}. A biofilm comprises a region with a high local density of bacteria that are immobilized on a wall or similar support. Biofilms are a widespread problem in health and technology, arising for instance in bacterial fouling of water pipes \cite{fouling}, and lethal infections in patients with cardiac valve implants \cite{valves}.
Biofilm formation generally involves chemical communication between individual cells, but the effect of this may still be representable in part as a density-dependent swim speed $v(\rho)$. 
Alternatively it is possible to connect the quorum sensing apparatus of bacteria to their self-propulsion mechanism, thereby creating a decreasing $v(\rho)$ directly by genetic engineering \cite{Hwa1,Hwa2}. MIPS might also occur by various non-biochemical means, such as simple crowding, hydrodynamically mediated surface accumulation \cite{RupertPRL}, or by secretion of viscosity-enhancing polymers \cite{exopoly}. 

The MIPS scenario also impacts on the study of pattern formation in bacterial colonies started on an agar plate from a localized source. These were observed before the discovery of quorum sensing (which is a local response to the concentration of a secreted chemical \cite{quorum}) and attributed instead to long-range chemotactic interactions, which modulate the persistence time $\tau$ in an orientation-dependent manner \cite{rings,budrene,tyson,confirm}. As shown in \cite{PNAS} and discussed in
\cite{MikeBrenner}, a simpler explanation can be found by coupling a theory of MIPS to a logistic growth law. The latter describes the tendency of bacterial populations to move towards a stable `target' density at which cell death and cell division are in balance. If this target density lies within the two-phase region of the MIPS, then coexisting uniform phases are unstable to population change, whereas a uniform state at the target density
is unstable to phase separation. The result is a kind of micro-phase separation which leads to patterns similar to those observed experimentally in growing colonies \cite{PNAS}.

\subsection{Experiments on Synthetic Swimmers}
\label{synthswim}

A variety of self-phoretic colloid particles undergo self-propulsion, in the presence of a fuel supply such as dissolved hydrogen peroxide \cite{howse,sen}
or another source of stored internal energy \cite{Herminghaus}. In most cases the observed system is quasi-two dimensional, because such swimmers tend to accumulate at container walls. (Also most of the swimming mechanisms used would not be sustainable for long in $d=3$ without running out of fuel \cite{BrownPoon}.) Such studies often report clustering \cite{TheurkauffPRL,ButtinoniBechinger,PalacciScience}, perhaps caused in part by attractive interactions. Similar clusters are seen in bacterial systems with colloidal attractions induced by polymer \cite{ChantalPNAS}; in the self-phoretic context the attractions could instead arise kinetically through cross-particle responses to reagent and product gradients \cite{TheurkauffPRL,PalacciScience,SriramRamin}. At higher densities, bulk phase separation has been reported, and attributed to a MIPS-like mechanism \cite{ButtinoniBechinger}. 

Even if attractive interactions are also present, it seems plausible
that MIPS-related physics is implicated in the observation of stable
cluster phases \cite{TheurkauffPRL,PalacciScience,ChantalPNAS}. In
these cases, partial phase separation occurs but seemingly gets
arrested after the formation of clumps of modest size. Such clusters
can be interpreted in kinetic terms as the consequence of the mutual
stalling of two particles in head-on collision, which then present an
obstacle that causes other particles to stall when they hit
it~\cite{RednerPRL}. But this is also the kinetic interpretation of
the MIPS mechanism: as such, it is far from obvious why this process
should ever be self-limiting. (Recall that in passive phase
separation, without long-range repulsive interactions, two
small phase-separated droplets can always lower their interfacial
energy by merging.) One idea is that weak activity could oppose,
rather than enhance, a passive tendency to phase separate, creating
motion that will break droplets apart~\cite{ChantalPRL}. Indeed, this
effect is already seen in purely repulsive ABPs, and is one of the
reasons why very large system sizes are needed to reliably distinguish
bulk phase separation from steady state density fluctuations at
intermediate length scales \cite{StenhammarPRL}.

In summary, there is some evidence for MIPS in synthetic colloidal
swimmers, although much more would be welcome. In any case there is an
unsolved mystery concerning the apparently widespread formation of
finite clusters that, unlike their counterparts in passive systems,
fail to achieve full phase separation. 

One possible explanation for the absence of fully-fledged MIPS in ABP experiments involves hydrodynamic interactions.  At low densities, these cause flocking and structure formation  \cite{Pedley, Ignacio}, but at high
density hydrodynamic interactions tend to suppress MIPS
\cite{Fielding,StarkSquirm}.  This has been explained as
follows. To model collisional slowing down with a theory in which $v$
depends on a coarse-grained density $\rho$, one arguably
needs particles to undergo several collisions before changing
direction, so that the density is sampled, and an average $v$ can be
defined, at scales below $\ell$. This ceases to hold at high density, because hydrodynamic torques cause particles to undergo a large rotation
each time they meet \cite{Fielding}. Though certainly important in
$d=2$ \cite{Fielding,StarkSquirm}, the strength of this effect in
$d=3$ is so far unknown. In three dimensions then, hydrodynamic interactions could be responsible for the arrest of coarsening or, if we are unlucky, they could destroy MIPS
entirely.

\section{BEYOND THE LOCAL APPROXIMATION}
\label{Sec8}
In this section we review more recent work that takes the theoretical picture of MIPS beyond the local approximation. This step is essential to fully understand the {\em dynamics} of phase separation, which is driven, in passive systems, by interfacial tension. As mentioned already, within the local approximation there are no gradient terms in the free energy functional, (given by \eqref{localone} and \eqref{localtwo}) so that the interfacial tension of the equivalent passive system is strictly zero \cite{Bray}. Hence the {\em dynamics} of MIPS entirely depends on terms so far neglected. 

\subsection{Phenomenological Approach}
\label{sec:GradientinF}
A natural phenomenological ansatz, which is also suggested by some types of simplified kinetic theory \cite{SpeckLowen}, is to add a square gradient term with constant coefficient $\kappa$ to the free-energy functional of the equivalent passive system:
\begin{equation}
{\cal F} = \int [f(\rho) + \frac{\kappa}{2}(\nabla\rho)^2] d^d\r \label{localthree} 
\end{equation}
where $f(\rho)$ obeys \eqref{localtwo}. Hence
\begin{equation}
\mu = \delta{\cal F}/\delta\rho = \ln \rho + \ln v(\rho) -\kappa \nabla^2\rho \label{much}
\end{equation}
This maintains the mapping between MIPS and passive phase
separation~\cite{TCPRL}, and leads to the well known
Cahn-Hilliard-Cook equation~\cite{Bray}:
\begin{eqnarray}
\dot\rho &=& -\nabla.{\bf J}\label{rhofirstbis}\\
{\bf J} &=& - M\nabla\mu + \sqrt{2M}{\bf \Lambda} \label{phenomenal}
\end{eqnarray}
where $M = D\rho$ is called the collective diffusivity. This approach has the great benefit of simplicity. In combination with a logistic population growth, it was used successfully in \cite{PNAS} to address patterning in bacterial colonies. It also allows a large body of knowledge on the passive case, for instance concerning critical behavior, to be adopted {\em en masse}.

\subsection{Continuum Model: Nonintegrable Gradient Terms}
\label{continuum}
A more general study requires a systematic gradient expansion of which the local approximation is the zeroth order term. In this context
there is no reason to expect the corrections to obey the
integrability condition, \eqref{integrable}, as was noted in
\cite{TCPRL}. Accordingly one should expect the mapping between MIPS
and passive phase separation to break down at this level.  The
simplest approach \cite{StenhammarPRL} is to ignore any gradient
contributions arising from the nonlocality of $M(\r) =
D([\rho],\r)\rho$ within \eqref{phenomenal}, but study systematically
those arising from the nonequilibrium chemical potential, $\mu =
\mu_{\rm id}+\mu_{\rm ex}$. Here $\mu_{\rm ex}$ is still defined via
Eq.~\ref{integrable} as $\mu_{\rm ex}(\r) \equiv \ln v([\rho],\r)$ but
this can no longer in general be written as a functional derivative
$\delta{\cal F}_{\rm ex}/\delta \rho$. In contrast to this, a
nonlocal diffusivity $M$ does not destroy integrability and, unless something else does, has
little effect on kinetics (with a few exceptions
\cite{BrayD}). 

Focusing therefore on $\mu_{\rm ex}$, the approach of \cite{StenhammarPRL} is to assume
\begin{equation}
\mu_{\rm ex}(\r)  = \ln v(\hat\rho(\r)) \label{rhohat}
\end{equation}
where $\hat\rho(\r)$ is a smeared density found by convolution of $\rho$ with an isotropic local kernel whose range is comparable to the persistence length $\ell = v(\rho)\tau$. This is the length scale on which one particle samples the density of its neighbors before changing orientation. Importantly, this range is itself density-dependent, at least for ABPs on which we focus here. When $\rho$ is slowly varying we have $\hat\rho = \rho + \gamma^2\nabla^2\rho$ with $\gamma = \gamma_0v(\rho)\tau$ and $\gamma_0$ of order unity.
Further expanding \eqref{rhohat} in gradients then gives
\begin{equation}
\mu  = \ln\rho + \ln v(\rho) - \kappa(\rho) \nabla^2\rho \label{mufinal}
\end{equation}
where $\kappa(\rho) = -\gamma_0^2\tau^2 v(\rho)v'(\rho)$. Since $\kappa$ is not constant, this form of $\mu$ is nonintegrable. One can however define a `nearest integrable model' as
\begin{equation}
{\cal F} = \int \left[f(\rho) +\frac{\kappa(\rho)}{2}(\nabla\rho)^2\right] d^d\r
\end{equation}
for which the chemical potential instead reads
\begin{equation}
\mu_{DB}  = \ln\rho + \ln v(\rho) - \kappa(\rho) \nabla^2\rho -\frac{\kappa'(\rho)}{2}(\nabla\rho)^2\label{mudb}
\end{equation}
The last term is an inevitable partner to a density-dependent
$\kappa$ coefficient in any system that obeys the principle of
detailed balance. Its absence for MIPS has interesting consequences
\cite{StenhammarPRL,Wittkowski} that we described in the next section.

\begin{figure}
\includegraphics[width=5in]{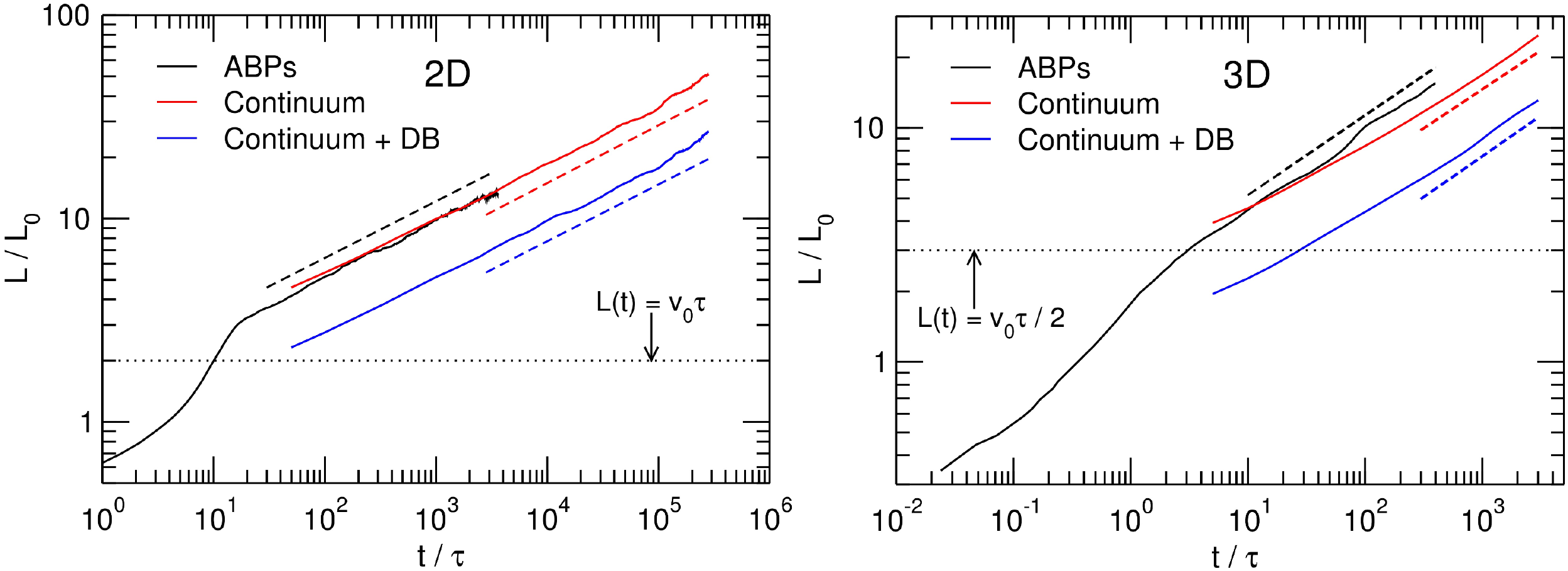}
\caption{(Left) Coarsening curves for ABPs, the continuum model
  violating detailed balance, and (shifted downwards for clarity) the
  nearest integrable model in $d=2$. (Right) the same in $d=3$. Dashed
  lines correspond to exponent $\simeq 0.28$ in $d=2$ and $\simeq
  0.33$ in $d=3$. Adapted from \cite{StenhammarSoft}, courtesy of
  J. Stenhammar.}\label{stenfigure}
\end{figure}

Equations \ref{rhofirstbis}, \ref{phenomenal} and \ref{mufinal}
comprise an explicit continuum model for MIPS whose only apparent
input (modulo the order unity factor $\gamma_0$) is the chosen
function $v(\rho)$. For ABPs, this is available from \eqref{linear}
but, as previously mentioned, a correction term must also be added to
$f(\rho)$ to prevent the density surpassing the close-packed
limit, $\rho^*$. The resulting theory can be compared with both direct
ABP simulations and the `nearest integrable model' (comprising
Eqs. \ref{rhofirstbis}, \ref{phenomenal} and \ref{mudb}). Figure
\ref{stenfigure} shows $L(t)$ curves for all three cases in $d=2,3$,
with apparent scaling exponents in $d=2$ somewhat below the value of
$1/3$ expected for diffusive coarsening in passive systems
\cite{StenhammarPRL,StenhammarSoft}. However, this shift is seen for
the nearest integrable model, as well as for the active continuum
model with DB violations \cite{StenhammarPRL}. Suggestively, passive
coarsening is known to show an altered exponent in the case where $M$
vanishes in the dense phase \cite{BrayD}, and so could give misleading
corrections to the asymptotic scaling when, as in ABPs, the
diffusivity is very small there.

More generally, the continuum model using \eqref{mufinal} gives a reasonably good account of domain shapes and dynamics when compared with direct ABP simulations. However, local fluctuations that violate time-reversal symmetry are under-represented in the continuum model \cite{StenhammarPRL,StenhammarSoft,Gompper}, for reasons that are not yet understood. Such fluctuations are prohibited altogether in the `nearest integrable model'. The fact that this prohibition has little effect on $L(t)$ shows these DB-violating fluctuations to be subdominant, at least in determining the rate of domain growth.

\subsection{Anomalous Phase Behavior: Active Model B}
A surprising numerical observation made in \cite{StenhammarPRL} is that the non-integrable gradient terms, while they have little consequence for coarsening dynamics, do affect the densities ($\rho_1$, $\rho_2$) of the coexisting phases. At first sight this is odd, because the common tangent construction makes no mention of any gradient terms. 
However, it does implicitly assume thermodynamic equilibrium and hence DB. There are other instances in physics where nonintegrable gradient terms alter an equilibrium result that appears not to involve gradients \cite{olmsted,watson}. 

The corresponding implications for MIPS were explored in \cite{Wittkowski} using a non-integrable generalization of `Model B'. The latter is a canonical model for diffusive phase separation in passive systems \cite{Hohenberg}. It introduces an order parameter $\phi$ which is, in this section, not the volume fraction but a linear transform of the particle density $\rho$
chosen so that the coexistence densities are at $\phi_{1,2}= \pm 1$.
Model B then writes a free energy functional
\begin{equation}
{\cal F} = \int\left[-\frac{\phi^2}{2} + \frac{\phi^4}{4}+\frac{\kappa}{2}(\nabla\phi)^2\right] d^d\r 
\end{equation}
with constant $\kappa$. The chemical potential is 
\begin{equation}
\mu = \delta{\cal F}/\delta\phi = -\phi + \phi^3 - \nabla^2\phi
\label{muMB}
\end{equation}
where we have set $\kappa = 1$ without loss of generality (this amounts to a rescaling of length). Suppressing any $\phi$-dependence of $M$ and choosing time units so that $M = 1$ gives
\begin{equation}
\dot\phi = -\nabla^2\mu + \nabla.{\bf \Lambda} \label{MB}
\end{equation}
\eqref{muMB} and \eqref{MB} comprise Model B for the purposes of passive phase separation studies.  The model gives $L(t)\sim t^{1/3}$, and captures other dynamical features such as the nucleation and growth kinetics in the regions of positive $f''$ within the common-tangent binodals. 

Active Model B \cite{Wittkowski} replaces \eqref{muMB} with
\begin{equation}
\mu = -\phi + \phi^3 - \nabla^2\phi +\lambda (\nabla\phi)^2
\label{muAMB}
\end{equation}
where the $\lambda$ term renders the model non-integrable, and can be viewed as the `distance' to the nearest integrable model (namely, Passive Model B). This structure can be compared with \eqref{mufinal} and \eqref{mudb} for ABPs. In the latter case, $\kappa(\rho)$ depends on density but a term in $\kappa'(\rho)(\nabla\rho)^2$ is missing from $\mu$. In Active Model B, $\kappa=1$ is constant but a term in $(\nabla \phi)^2$ is added. For simplicity this has a constant coefficient $\lambda$ which can have either sign but is negative for ABPs \cite{Wittkowski}. What matters is the {\em mismatch} between the $\nabla^2\phi$ and $(\nabla\phi)^2$ terms in the expression for $\mu$;  Active Model B captures this in the simplest possible way.

The model is simple enough to make analytical progress \cite{Wittkowski}. One finds that the common tangent construction is replaced by an `uncommon tangent' in which the two coexisting phases have the same chemical potential $\mu$ but unequal values of $\phi\mu-f$ which is, in thermodynamic language, the pressure. The reason for a shift in the coexistence conditions is, in this language, a discontinuity in pressure across the interface. Note that $\nabla\phi$ is not negligible in the interfacial region; indeed it is responsible (via $\kappa$) for the interfacial tension. The $\lambda$ term supplements this by a jump in thermodynamic pressure, which is linear in $\lambda$ for small values, but saturates at large ones in such a way that the coexisting densities can approach, but not enter, the spinodal region \cite{Wittkowski}. This kind of behavior is not possible in equilibrium, and shows that the DB violations that underly MIPS cannot be transformed away entirely. Nonetheless the corrections are weaker than they might have been. For instance, once DB is violated there is no guarantee that the densities of coexisting phases stay constant when the overall mean density in the system is changed. However, the uncommon tangent construction does preserve this feature.

\begin{figure}
\includegraphics[width=4.0in]{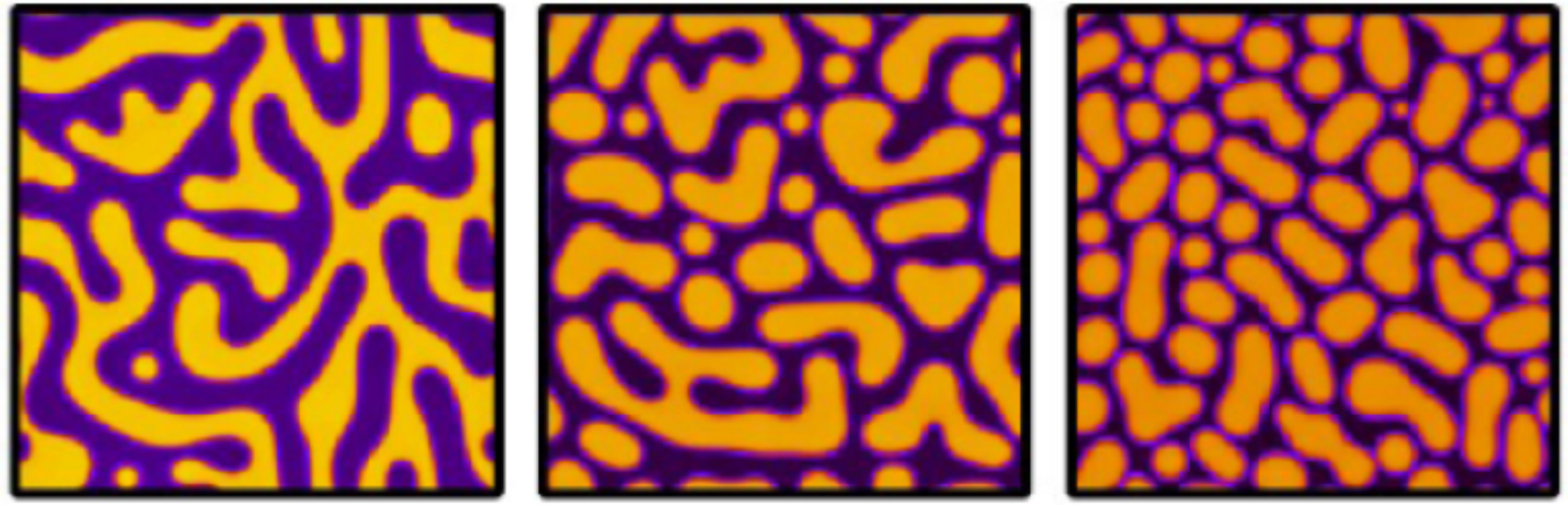}
\caption{Domain morphology in Active Model B during coarsening in $d=2$, for $\phi=\phi_0 = 0$ and $\lambda = 0,-1,-2$ (left, centre and right respectively). 
Courtesy of R. Wittkowski; adapted from \cite{Wittkowski}.}\label{Raphael}
\end{figure}

Active Model B explains the deviation from the common-tangent construction that was seen for ABP simulations \cite{StenhammarPRL}. However, a careful study of its coarsening behavior gives predictions qualitatively unaltered from the passive case, although numerically there is once again evidence of a reduced exponent for the temporal scaling of the domain size $L(t)$.  Figure \ref{Raphael} shows snapshots for various $\lambda$ taken during the coarsening process. Notably, Active Model B does not exhibit saturation of $L(t)$ at length scales smaller than the system size, and therefore cannot explain the existence of cluster phases in which coarsening has been reported to arrest at an intermediate length scale \cite{TheurkauffPRL,ButtinoniBechinger,PalacciScience,ChantalPNAS}. 

A tempting way forward from Active Model B is to construct an Active Model H, in which the $\phi$ field is coupled to a momentum-conserving solvent. For passive systems, the path from Model B to Model H is clear \cite{Hohenberg}, but this involves a thermodynamic relationship connecting the mechanical stress to a free energy derivative which breaks down for active matter.  At the time of writing, Active Model H therefore remains under development~\cite{Wittkowski}.

\section{CONCLUSION}
\label{Sec9}

In this review we have focused on one specific aspect of active matter
physics: the ability of motile particles, with isotropic interactions
whose only or main effect is to slow their propulsion speed at high
density, to undergo phase separation. By ignoring the complexities
presented by many real types of active matter, particularly
orientational interactions, workers on this topic have been able to
create a fairly detailed theoretical and numerical understanding of
the underlying physics. Large parts of this can be understood in terms
of an equivalent system of passive particles with attractive
interactions. This is interesting since it allows a highly developed
area of near-equilibrium statistical mechanics to be deployed in
modeling this specific class of active systems.

However, the equivalence is not complete, and its breakdown is
also interesting. One finds by simulation that, although much of the
behavior familiar in passive phase separation is retained, some
aspects of motility-induced phase separation (MIPS) irreducibly
violate detailed balance at mesoscopic and even macroscopic
scales. Indeed, during phase separation, on a mesoscopic scale within
the growing domains, particle currents arise that plainly violate time
reversal symmetry on that length scale. For active Brownian particles
these include lava-lamp type motion, and/or bubbles of
the minority phase forming continuously within a domain and then
moving to its surface \cite{StenhammarSoft,Gompper}. Macroscopically, one
observes modest but clear deviations in the densities of coexisting
phases from those predicted by globally minimizing an effective free
energy.  These features can be reproduced in part by simplified
continuum models in which detailed balance is violated by gradient
terms only.

Possibly because they address
only `active simple fluids' (without 
orientational interactions) the connection between theories of MIPS reviewed here and experiment remain somewhat tenuous at present. However,
phenomena resembling phase separation are certainly seen in some types
of experiments on bacteria and on synthetic colloidal swimmers. In the
latter case, there seems to be a generic tendency to form `cluster
phases' in which phase separation arrests at a
finite domain size. It is not yet clear whether this is a modified
form of MIPS (perhaps with passive attractions causing the arrest) or
a modified form of passive, attraction-driven phase separation
(perhaps arrested by activity).

This conundrum highlights a current deficiency of the theory: so far
we do not have a framework to combine MIPS-type effective attractions
with standard (i.e., passive) colloidal interactions inside a single
set of equations. The same basic obstacle arises whenever MIPS-like physics
is coupled to other phenomena: orientational
interactions, mixtures of active and passive particles, and
hydrodynamic forces. Such couplings are also very
challenging to address numerically, but one or more of them is present
in most experimental cases.  In the absence of decisive experimental
evidence for MIPS despite extensive numerical evidence, it may yet turn
out that, as far as experimental systems of ABPs are concerned, MIPS
is the dog that did not bark.\footnote{ Inspector Gregory: ``Is
  there any other point to which you would wish to draw my
  attention?''  \\ Holmes: ``To the curious incident of the dog in the
  night-time."  \\ Gregory: ``The dog did nothing in the night-time."
  \\ Holmes: ``That was the curious incident."  \cite{Sherlock}} Its
basic mechanism is extremely simple: a positive feedback between
slowing-induced accumulation and accumulation-induced slowing. When
both features are present, absence of MIPS may itself impart useful
mechanistic information about what else is happening in the system.
For example, as mentioned in Sec.~\ref{synthswim}, hydrodynamic interactions could suppress MIPS by causing inter-particle torques  \cite{Fielding}. 

Interaction torques are of course also important, even without hydrodynamic interactions, for aspherical swimmers.
The full physics of orientational interactions in active rods requires
one to address non-scalar order parameters describing either polar
(vector) or nematic (tensor) order. Theories of orientational ordering
and its effects in active systems have been extensively developed in
recent years as reviewed in \cite{ASMRMP,SriramReview1}, both for
`dry' systems (without momentum conservation, such as vibrated
granular rods) and `wet' ones such as bacterial swarms in
solution. Even without condensation into an orientationally ordered
state, incipient rotational order can enhance the tendency to undergo
MIPS in systems where a density-dependent swim speed $v(\rho)$ is
combined with an additive translational diffusion $D_t$
\cite{Peruani}. Furthermore, when MIPS induces the formation of
dense droplets, aligning interactions may lead to the appearance of
local order, hence making the droplets self-propel~\cite{Farrell}. All
this should be relevant to the study of active Brownian rods or
dimers~\cite{Peruani1,ChantalPNAS,BrownianRods1,BrownianRods2,BrownianRods3,mccandlish,Gomp2} where
MIPS-like phenomena, such as aggregation and swarm formation have been
reported. Since interparticle torques convey vectorial information
between particles, this is an area where the
physics of the Vicsek model~\cite{Vicsek} (which give flocking via
alignment interactions at fixed $v$) and that of MIPS (which stems
solely from slowing of $v$ at high density) overlap. This could be a
fruitful area for future studies. 

Finally, we mention the literature on simulating discrete
non-spherical swimmers with hydrodynamic interactions
\cite{HydroRods1,HydroRods2,HydroRods3}. While some of the observed phenomena may well be
related to MIPS, the additional presence of both near-field alignment interactions and far-field hydrodynamics, makes
any connection difficult to establish without further research.

\section*{ACKNOWLEDGMENTS} 

The authors {thank} particularly
Joakim Stenhammar, and also
Rosalind Allen,
Richard Blythe,
{Fred Farrell,
Christina Marchetti,}
Davide Marenduzzo,
{Ignacio Pagonabarraga,}
Wilson Poon,
Adriano Tiribocchi,
Alasdair Thompson,
Raphael Wittkowski,
and
Chantal Valeriani,
for discussions and collaborations relevant to this article. MEC is funded by a Royal Society Research Professorship and EPSRC Grant EP/J007404. This research was supported in part by the National Science Foundation under Grant No. NSF PHY11-25915.


\begin{thebibliography}{100}

\bibitem{PNASGiardina} Ballerini M, Cabibbo N, Candelier R, Cavagna A, Cisbani E {et al.}, 2008. \textit{Proc. Natl. Acad. Sci. USA} {\bf 105}, 1232-37.

\bibitem{poisson} Katz Y, Tunstr\o{}m K, Ioannou CC, Huepe C,  Couzin ID. 2011, \textit{Proc. Natl. Acad. Sci. USA} \textbf{108}, 18720-25. 

\bibitem{Bausch} Schaller V, Weber C, Semmrich C, Frey E, Bausch AR, 2010. \textit{Nature} \textbf{467} 73-77.

\bibitem{Sumino} Sumino Y, Nagai KH, Shitaka Y, Tanaka D, Yoshikawa K
  \textit{et al.}, 2012. \textit{Nature} {\bf 483} 448-52.

\bibitem{TheurkauffPRL} Theurkauff I, Cottin-Bizonne C, Palacci J, Ybert C, Bocquet L, 2012. \textit{Phys. Rev. Lett.} \textbf{108}, 268303.

\bibitem{ButtinoniBechinger} Buttinoni I, Bialke J, Kummel F, Lowen H, Bechinger C, Speck T. 2013. {\em Phys. Rev. Lett.} {\bf 110}, 238301.

\bibitem{PalacciScience} Palacci J, Sacanna S, Steinberg AP, Pine DJ, Chaikin PM, 2013. \textit{Science} \textbf{339}, 936-40.

\bibitem{DB} Bricard A, Caussin JB, Desreumaux N, Dauchot O, Bartolo D, 2013. \textit{Nature} \textbf{503}, 95-98.

\bibitem{Vicsek} Vicsek T, Czir\'ok A, Ben-Jacob E, Cohen I, Shochet O, 1995. \textit{Phys. Rev. Lett.} \textbf{75}, 1226.

\bibitem{ABP_review} Romanczuk P, B\"ar M, Ebeling W, Lindner B, Schimansky-Geier L, 2012.  \textit{Europ. Phys. Jour. Special Topics} {\bf 202}, 1-162.

\bibitem{CatesRPP} Cates ME, 2012. \textit{Repts. Prog. Phys.} {\bf 75}, 042601.

\bibitem{ASMRMP}  Marchetti MC, Joanny JF, Ramaswamy S, Liverpool TB, Prost J, Rao M, Simha RA, 2013. \textit{Rev. Mod. Phys.} {\bf 85}, 1143.

\bibitem{fouling} Flemming HC, 2002. {\em App. Microbiol. and Biotech.} {\bf 59}, 629-640.

\bibitem{valves} Costerton JW, Montanaro L, Arciola CR, 2005. {\em Int. J. Artificial Organs} {\bf 28}, 1062-1068.

\bibitem{biofilms} Hall-Stoodley L, Costerton JW and Stoodley P, 2004. {\em Nat. Rev. Microbiol.} {\bf 2}, 95-108.

\bibitem{scaffolds} Lee KY, Mooney DJ,  2001. {\em Chem. Rev.} {\bf 101} 1869-1879.

\bibitem{MFCs} Logan BE, Regan JM, 2006. {\em Trends. Microbiol.} {\bf 14} 512-518.

\bibitem{goldstein2} Sokolov A, Aranson IS, Kessler JO and Goldstein RE, 2007. {\em Phys. Rev. Lett.} {\bf 98} 158102.

\bibitem{activenema} Ramaswamy S, Simha RA,  Toner J, 2003. \textit{EPL} {\bf 62}, 196.

\bibitem{Peruani1} Peruani F, Deutsch A and Baer M, 2006. {\em Phys. Rev. E} {\bf 74}, 030904(R).

\bibitem{activeising} Solon A, Tailleur J, 2013. \textit{Phys. Rev. Lett.} {\bf 111}, 078101.

\bibitem{cisneros} Cisneros LH, Cortez R, Dombrowski C, Goldstein RE, Kessler JO, 2007. {\em Expts. in Fluids} {\bf 43} 737-53.

\bibitem{JTEPL} Tailleur J, Cates ME, 2009. \textit{EPL} {\bf 86}, 600002.

\bibitem{Palacci2010} Palacci J, Cottin-Bizonne C, Ybert C,  Bocquet L. 2010, \textit{Phys. Rev. Lett.} {\bf 105}, 088304.

\bibitem{StarkPRL} Enculescu M, Stark H, 2011. \textit{Phys. Rev. Lett.} {\bf 107}, 058301.

\bibitem{Austin1} Galajda P, Keymer J, Chaikin P, Austin R, 2007. {\em J. Bacteriol.} {\bf 189} 1033.

\bibitem{wan} Wan MB, Reichhardt CO, Nussinov Z, Reichhardt C, 2008. \textit{Phys. Rev. Lett.} {\bf 101} 018102.

\bibitem{rotorpaper1} Di Leonardo R, Angelani L, Dell'Arciprete D, Ruocco G, Iebba V {\em et al}, 2010. {\em Proc. Nat. Acad. Sci. USA} {\bf 107} 9541-45.

\bibitem{rotorpaper2} Angelani L, Di Leonardo R and Giancarlo R, 2009. {\em Phys. Rev. Lett.} {\bf 102} 048104.

\bibitem{rotorpaper3} Sokolov A, Apodaca MM, Grzybowski BA, Aronson IS, 2010. {\em Proc. Nat. Acad. Sci. USA} {\bf 107} 969-974.

\bibitem{quorum} Miller MB, Bassler BL, 2001. {\em Ann. Rev. Microbiol.} {\bf 55} 165-199.

\bibitem{thompson} Thompson AG, Tailleur J, Cates ME, Blythe RA, 2011. {\em J. Stat. Mech.} P02029.

\bibitem{Fily} Fily Y, Marchetti MC, 2012. \textit{Phys. Rev. Lett.} {\bf 108}, 235702.

\bibitem{RednerPRL} Redner GS, Hagan MF, Baskaran A, 2013. \textit{Phys. Rev. Lett.} {\bf 110}, 055701.

\bibitem{BialkeLowen} Bialk\'e J, L\"owen H, Speck T, 2013. \textit{EPL} {\bf 103}, 30008.

\bibitem{Ludo} Levis D, Berthier L, 2014. arxiv:1403.3410.

\bibitem{TCPRL} Tailleur J, Cates ME, 2008. {\em Phys. Rev. Lett.} {\bf 100}, 218103.

\bibitem{SchnitzerPRE}  Schnitzer MJ, 1993. {\em Phys. Rev. E} {\bf 48}, 2553-68.

\bibitem{VK} van Kampen NG, 1988. \textit{J. Phys. Chem. Solids} {\bf 49}, 673-77.

\bibitem{SchnitzerBerg} Schnitzer MJ, Block SM, Berg HC, Purcell EM, 1990. {\em Symp. Soc. Gen. Microbiol.} {\bf 46}, 15-33.

\bibitem{BergBook} Berg HC, 2003. {\em E. coli in Motion} (Springer, New York)

\bibitem{Krapivsky} Krapivsky PL, Redner S, Ben-Naim E, 2010. \textit{A Kinetic View of Statistical Physics}. (Cambridge University Press).

\bibitem{EPL2} Cates ME, Tailleur J, 2013. \textit{EPL} {\bf 101}, 20010.

\bibitem{Dean} Dean DS, 1996. {\em J. Phys. A} {\bf 29}, L613-17

\bibitem{Ito} \O{}ksendal B, 2003. \textit{Stochastic Differential Equations: an Introduction with Applications}. (Springer Berlin Heidelberg).

\bibitem{Hwa1} Liu C, Fu X, Liu L, Ren X, Chau CK, Li S \textit{et al.}, 2011. \textit{Science} {\bf 334}, 238-241.

\bibitem{Hwa2} Fu X, Tang LH, Liu C, Huang JD, Hwa T \textit{et al.}, 2012. \textit{Phys. Rev. Lett.} {\bf 108}, 198102.

\bibitem{SpeckLowen} Speck T, Bialk\'e J, Menzel AM, L\"owen H. 2013.  arXiv:1312.7242.

\bibitem{spinodal} Onuki A, 2002. \textit{Phase Transition Dynamics} (Cambridge University Press).

\bibitem{Bray} Bray AJ, 2002. \textit{Adv. Phys.} {\bf 51}, 481.

\bibitem{HenkesSoft}  Fily Y, Henkes S, Marchetti MC, 2014. \textit{Soft Matter} {\bf 10}, 2132-40.

\bibitem{Sandow1994} Schütz G, Sandow S, 1994. \textit{Phys. Rev. E} {\bf 49}, 2726.

\bibitem{TailleurJPA} Tailleur J, Kurchan J, Lecomte V, 2008. \textit{J. Phys. A} {\bf 41}, 505001.

\bibitem{golestanianRTP} Soto R, Golestanian R, 2014. \textit{Phys. Rev. E} {\bf 89}, 012706.

\bibitem{StenhammarPRL} Stenhammar J, Tiribocchi A, Allen RJ, Marenduzzo D, Cates ME, 2013. \textit{Phys. Rev. Lett.}{\bf 111}, 145702.

\bibitem{StenhammarSoft} Stenhammar J, Marenduzzo D, Allen RJ, Cates ME, 2014. \textit{Soft Matter} {\bf 14}, 1489-99.

\bibitem{Gompper} Wysocki A, Winkler RG, Gompper G, 2014. \textit{EPL} {\bf 105}, 48004. 

\bibitem{BerthierPack} Berthier L, 2013. arXiv:1307.0704.

\bibitem{Henkes} Henkes S, Fily Y, Marchetti MC, 2011. \textit{Phys. Rev. E} {\bf 84}, 040301.

\bibitem{RednerPRE} Redner GS, Baskaran A, Hagan MF, 2013. \textit{Phys. Rev. E} {\bf 88}, 012305.

\bibitem{Wittkowski} Wittkowski R, Tiribocchi A, Stenhammar J, Allen RJ, Marenduzzo D, Cates ME, 2013. arXiv:1311.1256.

\bibitem{RupertPRL} Nash RW, Adhikari R, Tailleur J, Cates ME, 2010. {\em Phys. Rev. Lett.} {\bf 104}, 258101.

\bibitem{exopoly} Sutherland IW, 2001. {\em Microbiol. UK} {\bf 147}, 3-9.

\bibitem{rings} Budrene EO, Berg HC, 1991. {\em Nature} {\bf 349}, 630-3.

\bibitem{budrene} Woodward DE, Tyson R, Myerscough MR, Murray JD, Budrene EO, Berg HC, 1995. {\em Biophys. J.} {\bf 68}, 2181-89.

\bibitem{tyson} Tyson R, Lubkin SR, Murray JD, 1999. {\em Proc. R. Soc. Lond. Ser. B} {\bf 266}, 299-304.

\bibitem{confirm} Budrene EO, Berg HC, 1995. {\em Nature} {\bf 376}, 49-53.

\bibitem{PNAS} Cates ME, Marenduzzo D, Pagonabarraga I, Tailleur J, 2010. {\em Proc. Nat. Acad. Sci. USA} {\bf 107}, 11715-20.

\bibitem{MikeBrenner} Brenner MP, 2010. {\em Proc. Nat. Acad. Sci. USA} {\bf 107} 11653-65.

\bibitem{howse} Howse JR, Jones RAL, Ryan AJ, Gough T, Vafabakhsh R, Golestanian R, 2007. {\em Phys. Rev. Lett.} {\bf 99}, 048102.

\bibitem{sen} Ibele M, Mallouk TE, Sen A, 2009.  {\em Ang. Chem. Int. Edn.} {\bf  48} 3308-12.

\bibitem{Herminghaus} Thutupalli S, Seemann R, Herminghaus S, 2011.  {\em New J. Phys.} {\bf 13}, 073021;
Dreyfus R et al. 2005. \textit{Nature} {\bf 437} 862-865.

\bibitem{BrownPoon} Brown A, Poon WCK, 2014. \textit{Soft Matter} {\bf 10}, 4016-4027.

\bibitem{ChantalPNAS} Schwarz-Linek J, Valeriani C, Cacciuto A, Cates ME, Marenduzzo D \textit{et al.}, 2012. \textit{Proc. Nat. Acad. Sci. USA} {\bf 109} 4052-4057.

\bibitem{SriramRamin} Saha S, Golestanian R, Ramaswamy S, 2013. arXiv:1309.4947

\bibitem{ChantalPRL} Mognetti BM, Sari\'c A, Angioletti-Uberti S, Cacciuto A, Valeriani C, Frenkel D, 2013. \textit{Phys. Rev. Lett.} {\bf 111}, 245702.

\bibitem{Pedley} Ishikawa T, Pedley TJ, 2008. {\em Phys. Rev. Lett.} {\bf 100}, 088103.

\bibitem{Ignacio}  Llopis I, Pagonabarraga I, 2006. {\em EPL} {\bf 75}, 999-1005.

\bibitem{Fielding} Fielding SM, 2012. arXiv:1210.5464.

\bibitem{StarkSquirm} Pohl O, Stark H, 2014. arXiv:1403.4063.



\bibitem{BrayD} Bray AJ, Emmott CL, 1995. \textit{Phys. Rev. B} {\bf 52}, R685-R688. 

\bibitem{olmsted} Lu CYD, Olmsted PD, Ball RC, 2000. \textit{Phys. Rev. Lett.} {\bf 84} 642-645.

\bibitem{watson} Watson SJ, Norris SA, 2006. \textit{Phys. Rev. Lett.} {\bf 96}, 176103.

\bibitem{Hohenberg} Chaikin P, Lubensky TC, 1995. \textit{Principles of Condensed Matter Physics}. (Cambridge University Press.) 


\bibitem{Sherlock} Conan-Doyle, A., 1894. \textit{The Memoirs of Sherlock Holmes. Ch.1: Silver Blaze.} (George Newnes Ltd., London.)

\bibitem{SriramReview1} Ramaswamy S, 2010. \textit{Ann. Rev. Cond. Mat. Phys.} {\bf 1} 323-245.

\bibitem{Peruani} Barre J, Chetrite R, Muratori M, Peruani F, 2014.  arXiv:1403.2364

\bibitem{Farrell} Farrell FDC, Marchetti MC, Marenduzzo D, Tailleur J. 2012. \textit{Phys. Rev. Lett.} {\bf 108}, 248101.

\bibitem{BrownianRods1} Baskaran A, Marchetti MC, 2008. \textit{Phys. Rev.  Lett.} {\bf 101} 268101.
\bibitem{BrownianRods2}  
Ginelli F, Peruani F, Baer M, 
Chate H, 2010. \textit{Phys. Rev. Lett.} {\bf 104} 184502.
\bibitem{BrownianRods3} 
Wensink H, L\"owen H, 2012. \textit{J. Phys. Cond. Mat.} {\bf 24} 464130.

\bibitem{mccandlish} McCandlish SR, Baskaran A, Hagan MF,
  2012. \textit{Soft Matt.} {\bf 8} 2527-34.
  
\bibitem{Gomp2} Abkenar M, Marx K, Auth T, Gompper G, 2013.
\textit{Phys. Rev. E} {\bf 88} 062314.  

\bibitem{HydroRods1} 
Hernandez-Ortiz JP, Graham MD, 2005. \textit{Phys. Rev. Lett} {\bf 95} 204501.
\bibitem{HydroRods2} 
Saintillan D, Shelley MJ, 2007. \textit{Phys. Rev. Lett} {\bf 99} 058102.
\bibitem{HydroRods3} 
Saintillan D, Shelley MJ, 2008. \textit{Phys. Rev. Lett} {\bf 100} 178103.



\end{thebibliography}
\end{document}